\documentclass{aa}  

\usepackage{graphicx}
\usepackage{txfonts}
\usepackage[normalem]{ulem}

\newcommand{\chandra}{{\it Chandra}}
\newcommand{\xmm}{{\it XMM-Newton}}

\newcommand{\suzaku}{{\it Suzaku}}

\newcommand{\tbvarabs}{{\tt tbvarabs}}
\newcommand{\vapec}{{\tt vapec}}
\newcommand{\vnei}{{\tt vnei}}
\newcommand{\pow}{{\tt powerlaw}}
\newcommand{\nh}{{$N_{\mathrm{H}}$}}

\newcommand{\ha}{H$\alpha$}
\newcommand{\hi}{$\ion{H}{i}$}
\newcommand{\hii}{$\ion{H}{ii}$}
\newcommand{\sii}{[$\ion{S}{ii}$]}
\newcommand{\oiii}{[$\ion{O}{iii}$]}

\begin{document}

   \title{First studies of the diffuse X-ray emission in the Large Magellanic Cloud with eROSITA}

   \titlerunning{First eROSITA studies of diffuse X-ray emission in the LMC}

   \subtitle{}

   \author{Manami Sasaki
          \inst{1}
          \and
          Jonathan Knies\inst{1}
          \and
          Frank Haberl\inst{2}
          \and
          Chandreyee Maitra\inst{2}
          \and
          J\"urgen Kerp\inst{3}
          \and
          Andrei M. Bykov\inst{4}
          \and
          Konrad Dennerl\inst{2}
          \and
          Miroslav D. Filipovi\'c\inst{5}
          \and
          Michael Freyberg\inst{2}
          \and 
          B\"arbel S. Koribalski\inst{5,6}
          \and
          Sean Points\inst{7}
          \and 
          Lister Staveley-Smith\inst{8,9}
          }

   \institute{
Dr.\ Karl Remeis Observatory, Erlangen Centre for Astroparticle Physics, Friedrich-Alexander-Universit\"{a}t Erlangen-N\"{u}rnberg, Sternwartstra{\ss}e 7, 96049 Bamberg, Germany \email{manami.sasaki@fau.de}
\and
Max-Planck-Institut f\"{u}r extraterrestrische Physik, Gie{\ss}enbachstra{\ss}e 1, 85748 Garching, Germany
\and
Argelander-Institut f\"{u}r Astronomie, Universit\"{a}t Bonn, Auf dem H\"{u}gel 71, 53121 Bonn, Germany
\and
Ioffe Institute, 26 Politekhnicheskaya, St Petersburg 194021, Russia
\and
Western Sydney University, Locked Bag 1797, Penrith South DC, NSW 2751, Australia
\and
CSIRO Astronomy and Space Science, PO Box 76, Epping, NSW 1710, Australia
\and
Cerro Tololo Inter-American Observatory/NSF's NOIRLab, Casilla 603, La Serena, Chile
\and
International Centre for Radio Astronomy Research (ICRAR), M468, University of Western Australia, Crawley, WA 6009, Australia
\and
ARC Centre of Excellence for All Sky Astrophysics in 3 Dimensions (ASTRO 3D), Australia
             }

   \date{Received April 12, 2021; accepted MMMMM DD, YYYY}

 
  \abstract
   {In the first months after the launch in July 2019, eROSITA (extended Roentgen Survey with an Imaging Telescope Array) onboard Spektrum-Roentgen-Gamma (Spektr-RG, SRG) performed long-exposure observations in the regions around supernova (SN) 1987A and supernova remnant (SNR) N132D in the Large Magellanic Cloud (LMC).}   
   {We analyse the distribution and the spectrum of the diffuse X-ray emission in the observed fields to determine the physical properties of the hot phase of the interstellar medium (ISM).}
   {Spectral extraction regions were defined using the Voronoi tessellation method. The spectra are fitted with a combination of thermal and non-thermal emission models. The eROSITA data are complemented by newly derived column density maps for the Milky Way and the LMC, 888 MHz radio continuum map from the Australian Square Kilometer Array Pathfinder (ASKAP), and optical images of the Magellanic Cloud Emission Line Survey (MCELS).}
   {We detect significant emission from thermal plasma with $kT$ = 0.2 keV in all the regions. There is also an additional higher-temperature emission component from a plasma with $kT \approx$ 0.7 keV. The surface brightness of the latter component is one order of magnitude lower than that of the lower-temperature component. In addition, non-thermal X-ray emission is significantly detected in the superbubble 30 Dor C. The absorbing column density \nh\ in the LMC derived from the analysis of the X-ray spectra taken with eROSITA is consistent with the \nh\ obtained from the emission of the cold medium over the entire area. Neon abundance is enhanced in the regions in and around 30 Dor and SN 1987A, indicating that the ISM has been chemically enriched by the young stellar population. In the centre of 30 Dor, there are two bright extended X-ray sources, which coincide with the stellar cluster RMC 136 and the Wolf-Rayet stars RMC 139 and RMC 140. For both regions, the emission is best modelled with a high-temperature ($kT >$ 1 keV) non-equilibrium ionisation plasma emission and a non-thermal component with a photon index of $\Gamma = 1.3$. In addition, we detect an extended X-ray source at the position of the optical SNR candidate J0529-7004 with thermal emission and thus confirm its classification as an SNR.}
   {Using data from the early observations of the regions around SN 1987A and SNR N132D with eROSITA we confirm that there is thermal interstellar plasma in the entire observed field. eROSITA with its large field of view and high sensitivity at lower X-ray energies allows us for the first time to carry out a detailed study of the ISM at high energies consistently over a large region in the LMC. We thus measure the properties of the interstellar plasma and the distribution of non-thermal particles and derive the column density of the cold matter on the line of sight.}


   \maketitle
%
\section{Introduction}

The Large Magellanic Cloud (LMC) is the largest satellite galaxy of the Milky Way at a distance of 50~kpc \citep[e.g.,][]{2014AJ....147..122D}.
It is a gas-rich dwarf galaxy with an almost face-on, planar disk and many interesting asymmetric features like spiral arms, off-centered bar, or a shell-like structure \citep{1955AJ.....60..126D}. 
Older stellar populations dominate the LMC mass and are either smoothly distributed in the disk or form a homogeneous stellar halo \citep{2004A&A...423...97B,2006A&A...460..459B}.
Young stars, on the contrary, are mainly found in the spiral arms \citep{2019IAUS..344...66Y}.
The stellar populations indicate that the last major star formation in the Magellanic Clouds (MCs) occured 200$\pm$50 Myr ago \citep[e.g.,][and references therein]{2019A&A...628A..51J}.

Optical emission line and radio images reveal that the LMC has a large number of bright \hii\ regions with a wide distribution in the disk.  
The most distinct object among them is the giant \hii-region 30~Doradus (30~Dor), also called the Tarantula nebula. It is located north of the eastern end of the stellar bar of the LMC and is host to the massive super-star cluster RMC 136. 

The distribution of cold gas as seen in \hi\ emission \citep{1974ApJ...190..291M,2003ApJS..148..473K} also reveals interesting structures in the eastern part of the LMC. As found by \citet{1992A&A...263...41L}, there are two distinct \hi\ components in the LMC, the L- and D-components.
The D-component covers most of the LMC and is located in the plane of the galaxy disk. The L-component on the other hand is much more localised and has radial velocities that are $\sim$30 -- 60 km s$^{-1}$ lower than the D-component. 
The \hi\ gas distribution suggests that there was a close encounter between the LMC and the Small Magellanic Cloud (SMC) about 150 -- 200 Myr ago \citep{1990PASJ...42..505F}.
\citet{2017PASJ...69L...5F} revisited the \hi\ gas in the LMC and reported that the D- and L-components show a complementary distribution. 
Additionally, they found a third component, called the I-component, with radial velocities between the L- and D-component. The I-component was most likely formed by  interactions between the two major components. 
In addition, \citet{2017PASJ...69L...5F} found correlation between a large elongated complex of molecular clouds seen in CO called the CO ridge \citep{1999PASJ...51..745F,2001PASJ...53..971M,2008ApJS..178...56F} and the likely interaction region. The CO ridge is located west of 30~Doradus and extends towards the South.

At high energies, the LMC is known to be host to bright 
supernova remnants (SNR, e.g., N132D) and is a unique place in which we can study the evolution of a supernova (SN) into an SNR (SN 1987A).
South of 30~Dor, east of the CO ridge, a large diffuse structure was observed in X-rays in the ROSAT survey of the LMC, called the X-ray spur \citep{1997A&A...323..585B,2001ApJS..136...99P}.
The X-ray spur is located south of 30~Dor and the supergiant shell LMC 2 (LMC-SGS 2) and seems to coincide with the regions in which the L- and D-components of \hi\ collide with each other. The analysis of X-ray data taken with \xmm\ has shown that the X-ray spur was most likely caused by the \hi\ collision
\citep{2021A&A...648A..90K}. While in 30~Dor and in regions around it, the interstellar medium (ISM) was heated by the stellar winds and supernovae of massive stars, there are no indication of stellar heating in the X-ray spur.

In this paper, we present the analysis of the diffuse X-ray emission around 30~Dor and SNR N132D in the LMC based on the first observations with eROSITA  \citep[extended ROentgen survey with an Imaging Telescope Array,][]{2012arXiv1209.3114M,2021A&A...647A...1P}, taken during the early commissioning and calibration phase of the mission. 
eROSITA is a new X-ray telescope which was launched on July 13, 2019 onboard the  Spektrum-Roentgen-Gamma (Spektr-RG, SRG) spacecraft.
eROSITA consists of seven telescope modules (TMs), each equipped with a CCD detector. With its large field of view 
with a diameter of $\sim$1 degree
and high sensitivity in the energy band up to 10 keV, eROSITA allows us to study the distribution and the spectral properties of large extended X-ray emission for the first time.
We supplement our study with the latest generation of radio survey data from the Australian Square Kilometer Array Pathfinder \citep[ASKAP,][]{2008ExA....22..151J,2021PASA...38....9H}
and optical emission line images of the Magellanic Cloud Emission Line Survey \citep[MCELS,][]{2004AAS...20510108S}.
The studies of the point sources in the same observations and of SN 1987A are presented in separate papers by  Haberl et al. (in prep.) and Maitra et al. (in prep.), respectively.
We will report on the study of the diffuse X-ray emission in the entire LMC, which will be based on the eROSITA All-Sky Survey (eRASS) data in future publications.

\section{Data}

\subsection{eROSITA}

During the journey to Lagrange point L2, where SRG has been operating in a stable orbit ever since, commissioning tests and calibration of the instruments were carried out, along with performance verification observations. The first-light observation of eROSITA was pointed at SN 1987A next to 30~Dor, while several calibration observations have been performed around SNR N132D. 
We use the data from eROSITA observations in the LMC obtained during the commissioning and calibration phases 
(see Table \ref{obslst}).
The observations were carried out in the pointing or the field-scan mode and are roughly pointed at PSR J0540--6919, SN 1987A, and SNR N132D, some of them with offsets of 20\arcmin\ for calibration purposes. 
The data have been processed with the 
eROSITA Science Analysis Software System (\texttt{eSASS}, Brunner et al., 2021, submitted).
The eSASS pre-processing pipeline produces energy calibrated event files, which can then be used for the analysis. For the Early Data Release (EDR) we use the data of the processing version c001.

\begin{table*}
\centering
\caption{\label{obslst}
List of used eROSITA observations.
}
\begin{tabular}{llllll}
\hline
ID & Target & Observation Type & Instruments & Start Date & Pointing Exposure [ks]\\
\hline
700016 & SN 1987A & Comissioning & TM3, 4 & 15-09-20 & 21 \\
700161 & SN 1987A & First Light & TM1, 2, 3, 4, 5, 6, 7 & 18-10-20 & 80 \\
700156 & SNR N132D & Calibration & TM5, 6 ,7 & 10-10-20 & 48 \\
700179 & SNR N132D & Calibration & TM1, 2, 3, 4, 5, 6, 7 & 22-11-20 & 60 \\
700182 & SNR N132D & Calibration & TM1, 2, 3, 4, 5, 6, 7 & 27-11-20 & 40 \\
700183 & SNR N132D & Calibration & TM1, 2, 3, 4, 5, 6, 7 & 25-11-20 & 40 \\
700184 & SNR N132D & Calibration & TM1, 2, 3, 4, 5, 6, 7 & 23-11-20 & 40 \\
700185 & SNR N132D & Calibration & TM1, 2, 3, 4, 5, 6, 7 & 25-11-20 & 40 \\
700205 & PSR J0540--6919 & ART-XC Calibration & TM1, 2, 3, 4, 5, 6, 7 & 07-12-20 & 30 \\
700206 & PSR J0540--6919 & ART-XC Calibration & TM1, 2, 3, 4, 5, 6, 7 & 07-12-20 & 20 \\
700207 & PSR J0540--6919 & ART-XC Calibration & TM1, 2, 3, 4, 5, 6, 7 & 07-12-20 & 36 \\
\hline
\end{tabular}
\end{table*}

\begin{figure}
\centering
\includegraphics[width=.49\textwidth,trim=30 210 90 20, clip]{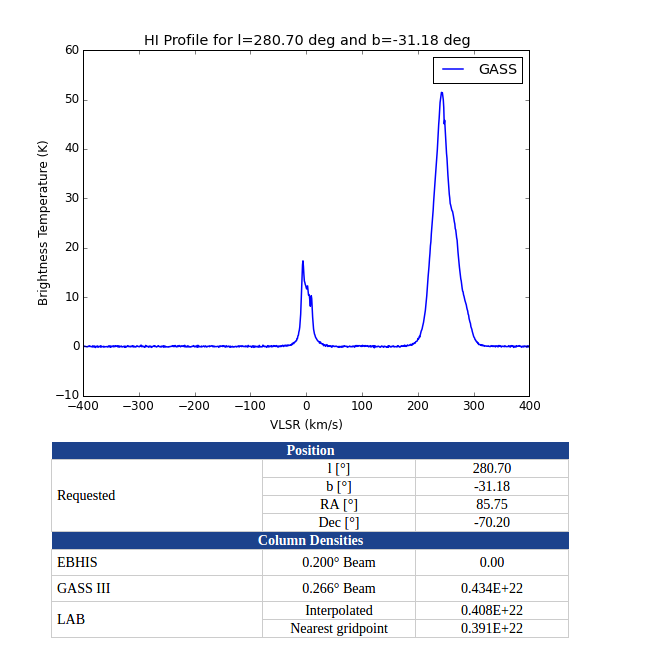}
\caption{\label{hiprof}
\hi\ velocity profile at an example position in the LMC
(RA = 85.75\degr, Dec = --70.20\degr).}
\end{figure}

\subsection{\nh\ Map}\label{nhcold}

X-rays interact via photoelectric absorption with baryonic matter distributed along the line of sight. The absorption cross section is a function of energy $\sigma \propto E^{-3}$ \citep[][]{2000ApJ...542..914W}, implying that the low energy portion of the total X-ray spectrum is strongly affected by photoelectric absorption. Hydrogen and helium are the most abundant elements in space. Their abundance ratio is still very close to that value defined by the primordial nucleosynthesis. Because of hydrogen's dominance in abundance, the number of atomic and molecular hydrogen is used as a proxy for the total number of interstellar and intergalactic gas atoms.

The amount of neutral atomic hydrogen can be measured via the \hi\ 21-cm line emission. Here, we make use of the HI4PI survey \citep[][]{2016A&A...594A.116H}\footnote{\url{https://www.astro.uni-bonn.de/hisurvey/AllSky_profiles/}}, which is based on the data of the Effelsberg-Bonn HI Survey \citep[EBHIS,][]{2016A&A...585A..41W} and the third revision of the Galactic All-Sky Survey  \citep[GASS,][]{2015A&A...578A..78K}. HI4PI quantifies the total \hi\ hydrogen atoms within the velocity range of $-600 \leq v_\mathrm{LSR}\mathrm{[km\,s^{-1}]} \leq 600$, comprising the full radial velocity range covered by the Milky Way \hi\ \citep[][]{2009ARA&A..47...27K} and the Magellanic Cloud System \citep[][]{2005A&A...432...45B}.
The integrated HI4PI $N_\mathrm{HI}$ map therefore displays a complex superposition of the \hi\ emission of the Milky Way and that of the Magellanic Clouds. Both however populate very different radial velocity ranges. The emission in the Magellanic Clouds is found at $200 \leq v_\mathrm{LSR}\mathrm{[km\,s^{-1}]} \leq 350$ \citep{2003ApJS..148..473K}, while the Milky Way's \hi\ emission is around $v_\mathrm{LSR}\mathrm{[km\,s^{-1}]}\simeq 0$ \citep[see Fig.\,\ref{hiprof};][]{2005A&A...432...45B}. We split up the \hi\ data into these two velocity regimes. 

\begin{figure}
\centering
\includegraphics[width=.49\textwidth]{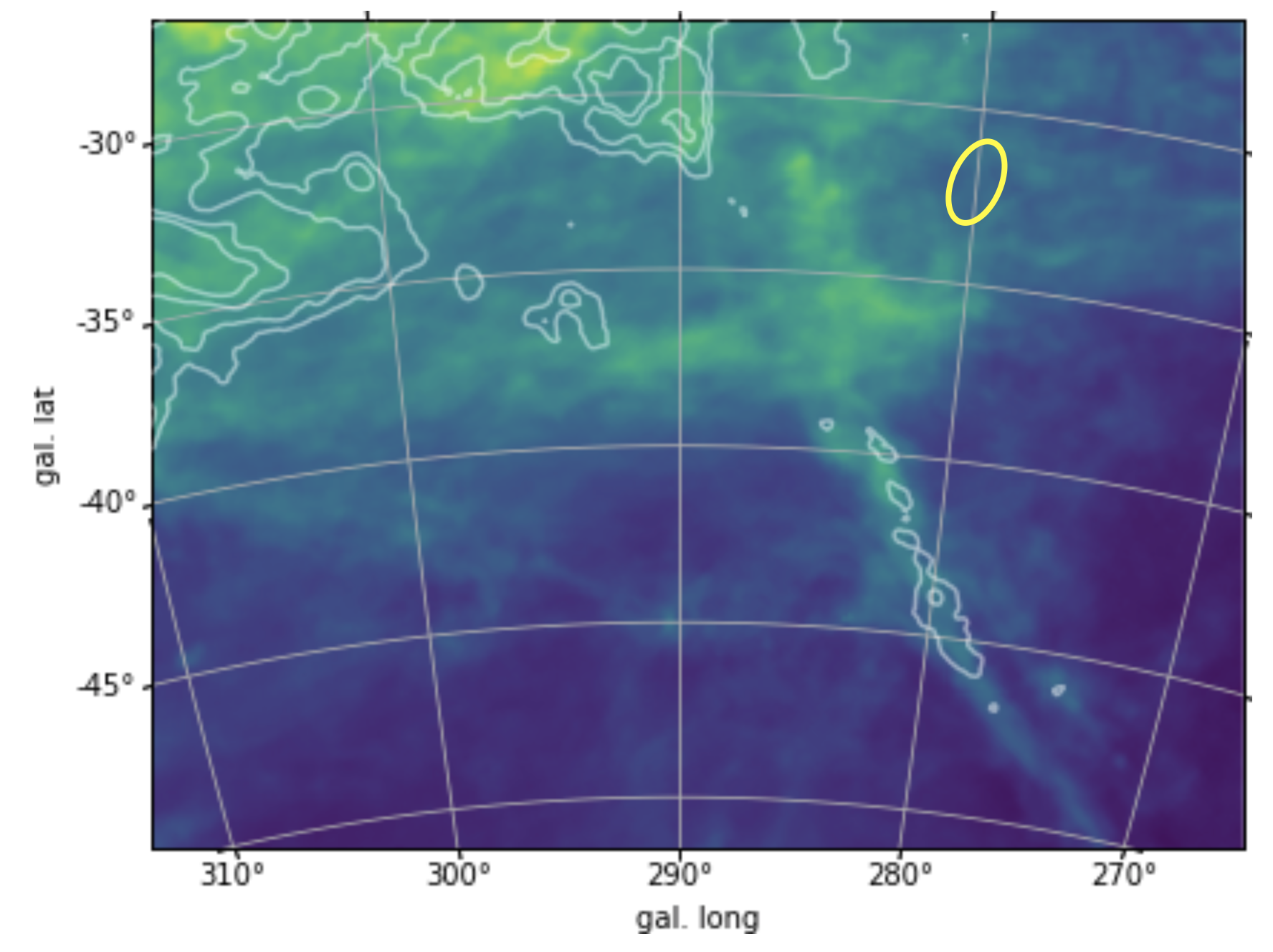}
\caption{\label{hih2ima}
Displayed in colour is the spatial distribution the Milky Way's $N_\mathrm{HI}$ column density 
for $V_\mathrm{LSR}$ =
--155 km/s to +80km/s
above a threshold of $N_\mathrm{HI} = 3 \times 10^{19}$ cm$^{-2}$ with a maximum of $1 \times 10^{21}$ cm$^{-2}$. Superimposed as contours is the 
spatial distribution of the Milky Way gas with excess optical extinction derived from the $N_\mathrm{HI}/A_\mathrm{V}$ ratio (see text for details).
The contours correspond to $ N_\mathrm{H_2} = 5 \times 10^{19}$ cm$^{-2}$ as well as to 1, 2 and $3 \times 10^{20}$ cm$^{-2}$. 
The LMC is located at $l = 280^\circ$, $b = -33^\circ$, the SMC at $l = 302^\circ$, $b = -44^\circ$.
The yellow ellipse marks the area which was observed with eROSITA.
}
\end{figure}


Observations of the 21-cm line does not take into account hydrogen nuclei 
in molecular phase.
Therefore, $N_\mathrm{HI}$ is only a lower limit to the true amount of hydrogen nuclei causing the soft X-ray absorption. To identify regions with significant amounts of molecular hydrogen we search for deviations from the median of the field-averaged gas-to-dust ratio $N_\mathrm{HI}/A_\mathrm{V}$.
Neither the number of dust grains nor the number of hydrogen nuclei is modified during a phase transition from \hi\ to $\mathrm{H_2}$ but the $N_\mathrm{HI}/A_\mathrm{V}$ ratio
is.
Regions with significantly low values for $N_\mathrm{HI}/A_\mathrm{V}$ spatially mark molecular gas.

With the aim of getting a reliable estimate of the foreground $N_\mathrm{H, Gal}$ because it modifies the soft X-ray emission from the LMC by photoelectric absorption,
we cross-correlate the HI4PI data with the interstellar reddening $E_\mathrm{B-V}$ data from \cite{1998ApJ...500..525S}. We apply the correction $E_\mathrm{B-V}(\mathrm{true}) = 0.884\cdot E_\mathrm{B-v}$ \citep{2011ApJ...737..103S}. According to \cite{1989ApJ...345..245C} and \cite{2001ApJ...548..296W} the visual extinction of the diffuse interstellar medium is $A_\mathrm{V} = 3.1 E_\mathrm{B-V}$, which is adopted in the following.
We evaluate $N_\mathrm{HI}/A_\mathrm{V}$ across an area of about 8000\,deg$^2$, which is sufficiently large to distinguish between the Magellanic Cloud System and the Milky Way halo gas.
For the Milky Way gas  we find a value of $N_\mathrm{HI}/A_\mathrm{V} = (2.04\pm0.15)\times 10^{21}\mathrm{cm^{-2}\,mag^{-1}}$, which is compatible with \cite{2009MNRAS.400.2050G} who determined $(2.21\pm0.09)\times 10^{21}\,\mathrm{cm^{-2}\,mag^{-1}}$. 
To obtain a map of the distribution of the excess extinction regions,
we calculated the mean
$N_\mathrm{HI}/A_\mathrm{V}$ for the Milky Way halo gas 
by applying a spatial mask to exclude the \hi\ emission of the Magellanic Cloud System.
Excess extinction regions, which deviate significantly from the field-averaged median value, are shown by the contour lines in Fig.\,\ref{hih2ima}.
While the \hi\ data allow to separate between the Milky Way and the Magellanic Cloud System emission, it is not possible to separate the optical extinction data. 
We do not find any evidence for additional neutral large scale structures in the immediate vicinity of our field observed with eROSITA, indicating that there are no significant amounts of molecular gas in the Milky Way Galaxy in this area of the sky.
The LMC X-ray radiation is  dominantly modulated only by its intrinsic \hi\ distribution (see Sect.\,\ref{absorption}).
On smaller angular scales additional soft X-ray shadows are expected because of the three dimensional structure of neutral 
and molecular
interstellar medium that belongs to the LMC itself.

\begin{figure*}
\centering
\raisebox{4.5mm}[0mm][0mm]{\includegraphics[width=.49\textwidth]{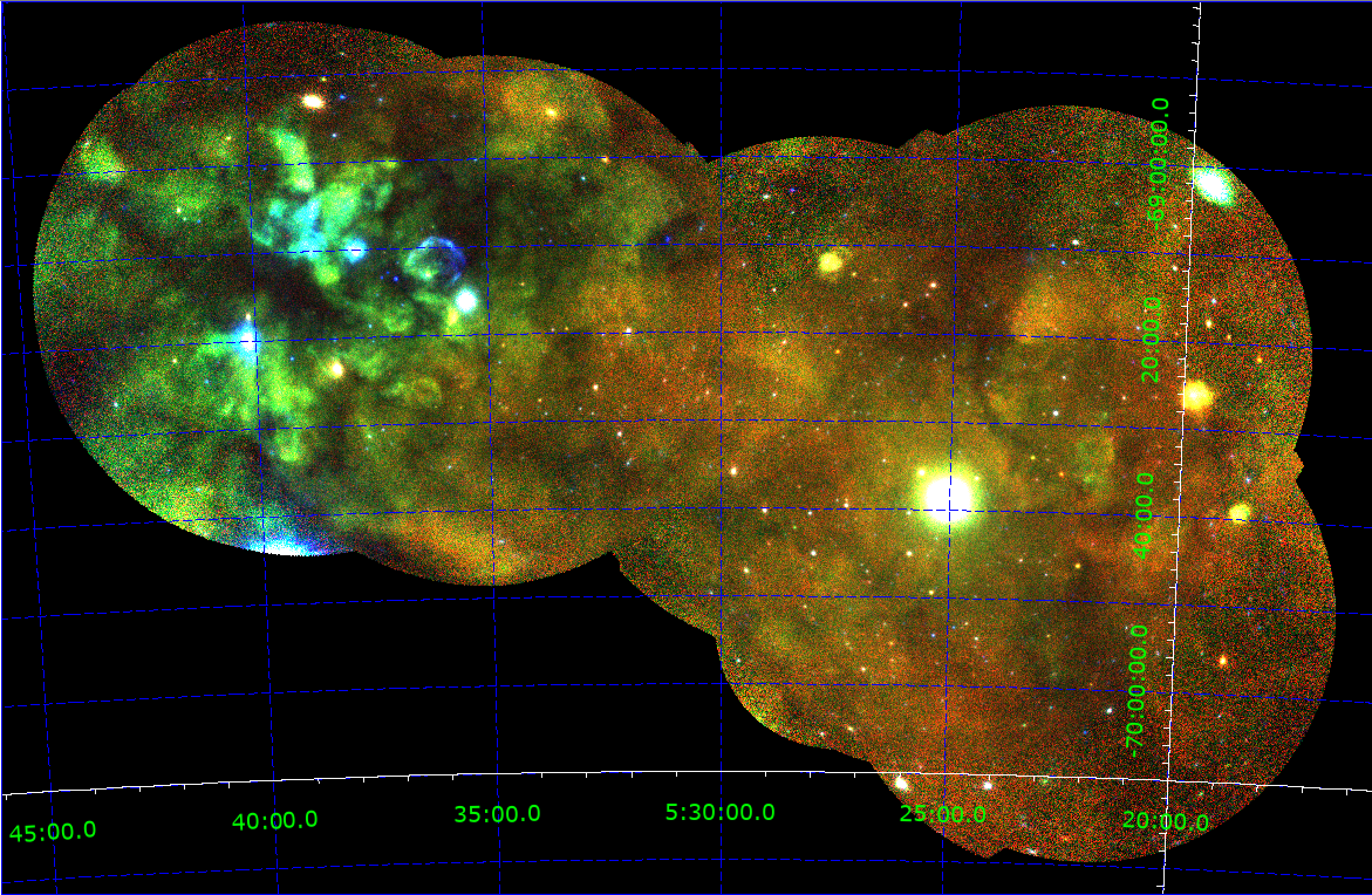}}
\includegraphics[width=.49\textwidth]{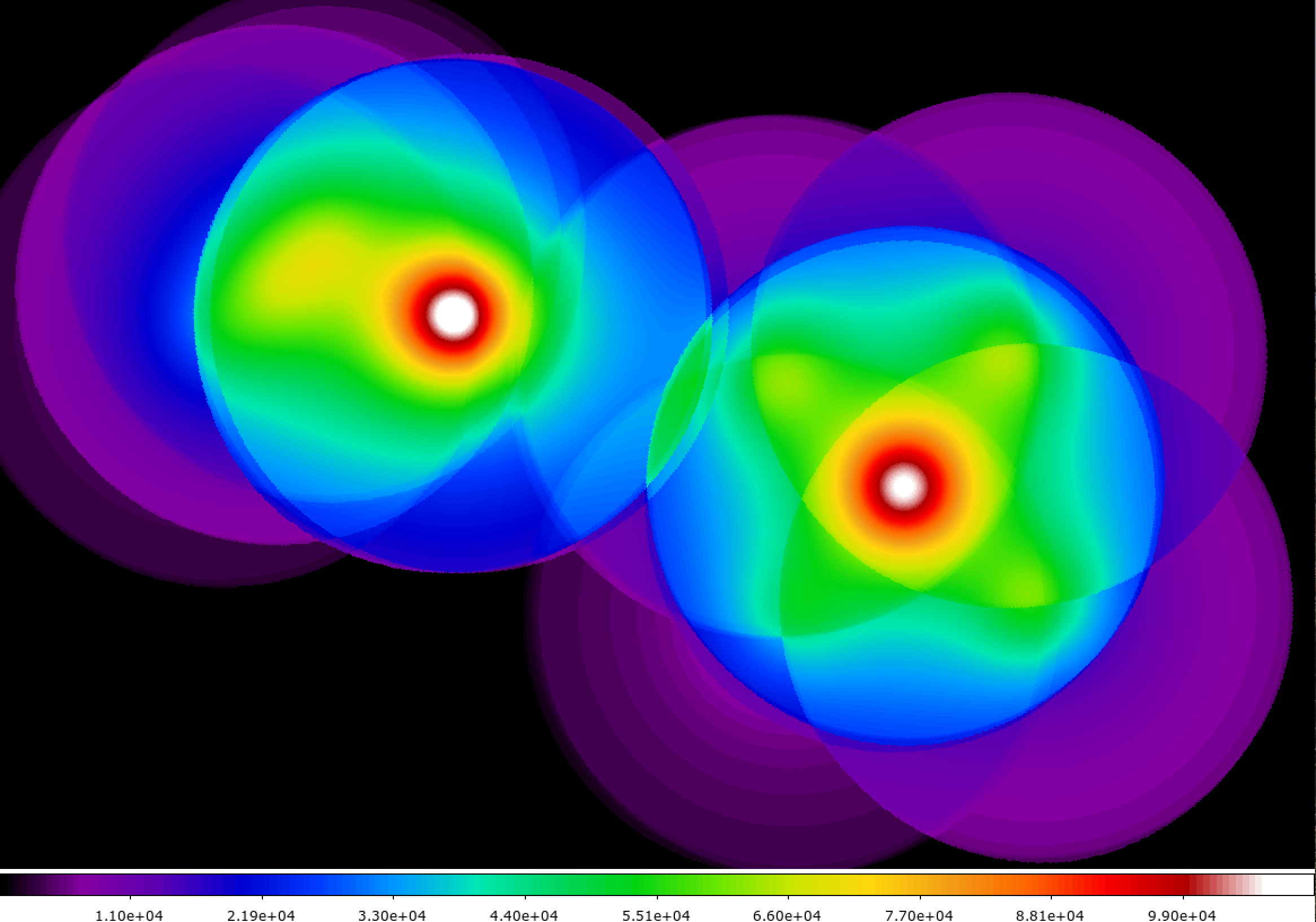}
\caption{\label{mosaic}
Left: Exposure-corrected mosaic image of the SN 1987A and SNR N132D regions
(red: 0.2 -- 0.5 keV, green = 0.5 -- 1.0 keV, blue = 1.0 -- 2.0 keV). Right: Mosaic of exposure maps of all observations in the entire energy range (0.2 -- 10.0 keV) shown in linear scale in the range of 0 ks to 110 ks.
}
\end{figure*}

\section{X-ray Analysis}

\subsection{Images}

eROSITA data were analysed using the \texttt{eSASS} version \texttt{eSASSusers\_201009}.
We created exposure-corrected mosaic images using the data in the broad 
band of 0.2 -- 10.0 keV and in the following sub-bands: 
0.2 -- 0.5 keV, 0.5 -- 1.0 keV, 1.0 -- 2.0 keV, and 2.0 -- 10.0 keV.
To create the images, event files for all seven telescope modules (TMs)
of eROSITA were binned with a 
bin size of 160 pixels, 
yielding counts images with a pixel size of 8\arcsec. These counts 
images and corresponding exposure maps were created for each observation 
and energy band.
After combining the images and exposure maps into large mosaics of all
observations, the mosaics of the counts images were divided by the
respective
exposure-map mosaics to create exposure-corrected mosaic images.
Figure \ref{mosaic} shows the mosaic images of all data in
Table \ref{obslst} below 2.0 keV and 
an exposure map of all observations in the broad band of 0.2 to 10.0 keV. For creating the mosaic images, we applied a lower cut for the exposure of 1 sigma below the mean value to mask out the outer areas of the FOVs in which the photon statistics were low. 

In Wolter type-I telescopes like eROSITA, single reflections of photons from nearby X-ray sources outside the field of view on the
second hyperboloid mirror shells 
can also reach the detectors.
Even though baffles in front of the telescopes can reduce the effect, it
will result in stray light in the data and contaminate both the images and the spectra. In our case, single reflections from photons from SNR N132D can become significant in observations to the
East.
We therefore carefully inspected the images for possible stray light,
which would have been visible as enhanced arc-like structure in one half of the field of view of each observation on the side close to N132D.
Fortunately, it caused no visible effect in the images. For the spectral analysis, for which we used the data of the observations 700156, 700161, and 700179, any contamination by additional emission from N132D will only be visible in the spectra extracted from 700161. As no enhancement in emission or change in spectral parameters was found in the spectral fits in regions that might be affected by stray light from N132D either, we assume that, also for the spectral analysis, the effect is negligible.

For the study of the diffuse emission, first, the standard eSASS source 
detection routine 
(Brunner et al., 2021, submitted)
was applied to all observations.
Source detection was performed in the 0.2 –- 2.3 keV band for each observations using all available TMs.
First, the sliding-box detection task \texttt{erbox} is run in local mode creating a preliminary list of sources, which is used to create a background map using the task \texttt{erbackmap}. 
Next, \texttt{erbox} is run in global mode using the background map.
This second source list is used to create an updated background map. These two files are then used as input for the task \texttt{emldet}, which carries out point-spread-function fitting to the sources, yielding a final source catalog with information such as the position, positional error, detection likelihood, etc.
In the next step, all point sources are removed from the 
event file of each observation.
We also manually defined regions for additional emission 
from very bright sources, which cause extended emission, and
stray light from 
the X-ray binary LMC X-1, which is located close the field of view and causes narrow arc-like features at the southern edge of the field of views in the eastern pointings, seen as blue emission in Fig.\,\ref{mosaic}.
These regions were also 
excluded from the data. 
Images were created from the cheesed event files for each observation in the same way as for the full mosaic.
In Fig.\,\ref{cheesed} we show the 
cheesed
mosaic images 
in the same energy bands as in Fig.\,\ref{mosaic}.

\subsection{Spectral Analysis}

\begin{figure}
\centering
\includegraphics[width=.49\textwidth]{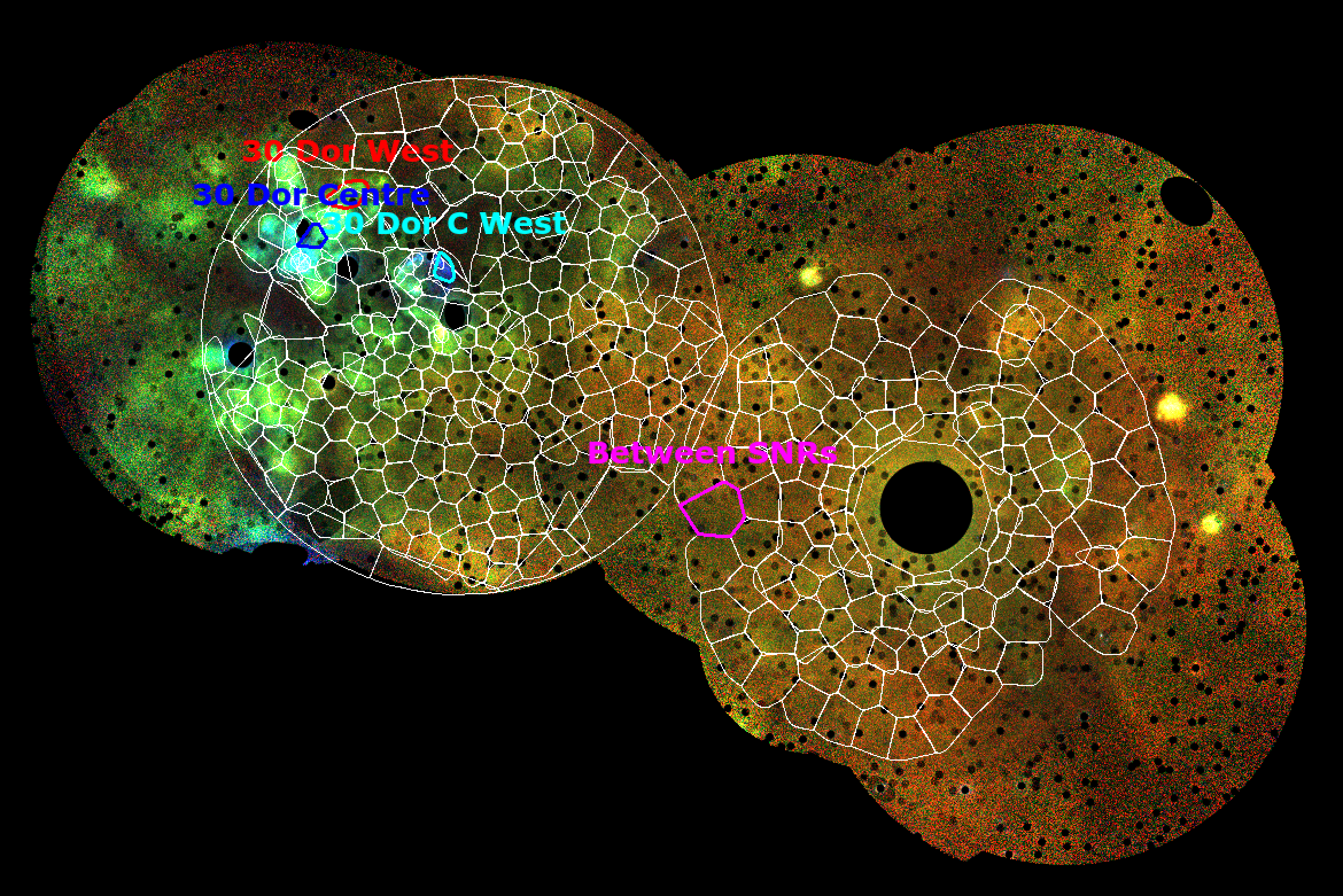}
\caption{\label{cheesed}
Exposure-corrected mosaic image of the SN 1987A and SNR N132D regions without
point sources
(red: 0.2 -- 0.5 keV, green = 0.5 -- 1.0 keV, blue = 1.0 -- 2.0 keV)
and regions used for spectral analysis.
The spectra of the regions marked in colour (30 Dor west, 30 Dor centre, 30 Dor C west, and between SNR N132D and SN 1987A in red, blue, cyan, and magenta, respectively) are shown in Fig.\,\ref{spectra}.
}
\end{figure}

Using the event files from which the point sources have been excluded,
we extracted spectra in regions that have been defined based on the
photon statistics using the Voronoi tessellation algorithm 
\citep{2003MNRAS.342..345C}.
For the spectral analysis, we focus on observations with SN 1987A and 
SNR N132D at or close to the aimpoints, which have high exposure ($>$20 ks,
700156, 700161, 700179).
Voronoi tessellation was not performed on the entire data of each observation at once,
since otherwise obviously continuous emission (e.g., in 30 Dor) was divided and merged with surrounding emission with lower brightness. Instead, we defined large regions based on similar surface brightness levels, each of which was tessellated.
In general, the diffuse emission is fainter in the western region around SNR N132D. We used a signal-to-noise ratio of
$>$50 for this region, which allows us to reach a good spatial resolution and at the same time yields spectra with good photon statistics. The region around SN 1987A shows brighter diffuse emission, which allowed us to define smaller extraction regions with a signal-to-noise ratio of $>$100 (see Fig.\,\ref{cheesed}).

The observational data are contaminated with particle-induced, non-X-ray 
background
and with astrophysical X-ray background \citep[for first studies of the
background measured with eROSITA, see][]{2020SPIE11444E..1OF}. 
For the spectral analysis, all background components were modelled and fit
simultaneously with the source components.
The particle background dominates the data at higher energies and
the higher energy band hence allows to determine its flux level.
Therefore, to estimate the particle background, we used the data up 
to 9.0 keV, even though no diffuse emission was observed above $\sim$7.0 keV.
The particle background consists of a continuum component that can be described with two to three power-law models with additional emission lines caused in the telescope.
The astrophysical X-ray background was modelled as a combination of emission 
from the Local Hot Bubble, Galactic halo, and the extragalactic X-ray 
background.
To estimate the astrophysical X-ray background, emission from non-source regions in the eROSITA EDR data from nearby observations  (pointed at the globular cluster 47 Tuc and the galaxy cluster A3158) were used. The spectral model parameters were determined by fitting the spectra of these non-source regions. For the analysis of the diffuse emission in the LMC, we fixed the model parameters for the astrophysical background and scaled it with a multiplicative constant parameter, which was free in the fit. 

\begin{figure}
\centering
\includegraphics[width=.49\textwidth,trim=0 3 3 0, clip]{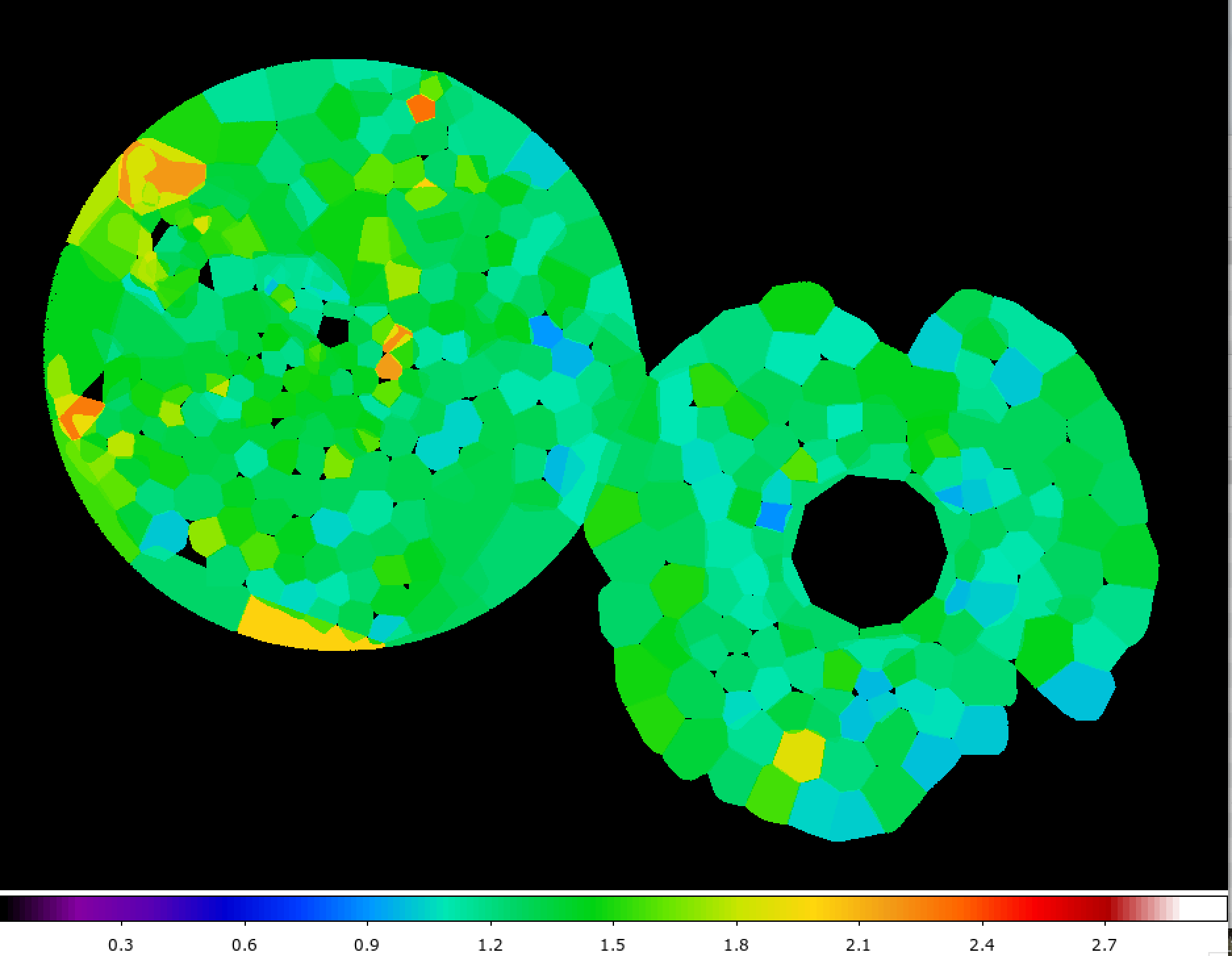}
\caption{\label{redchi}
Parameter map showing the values for $\chi^2$/d.o.f. of the best-fit model for the spectra of each region.
}
\end{figure}

\begin{figure*}[ht]
\includegraphics[width=.5\textwidth,trim=80 40 110 30,clip]{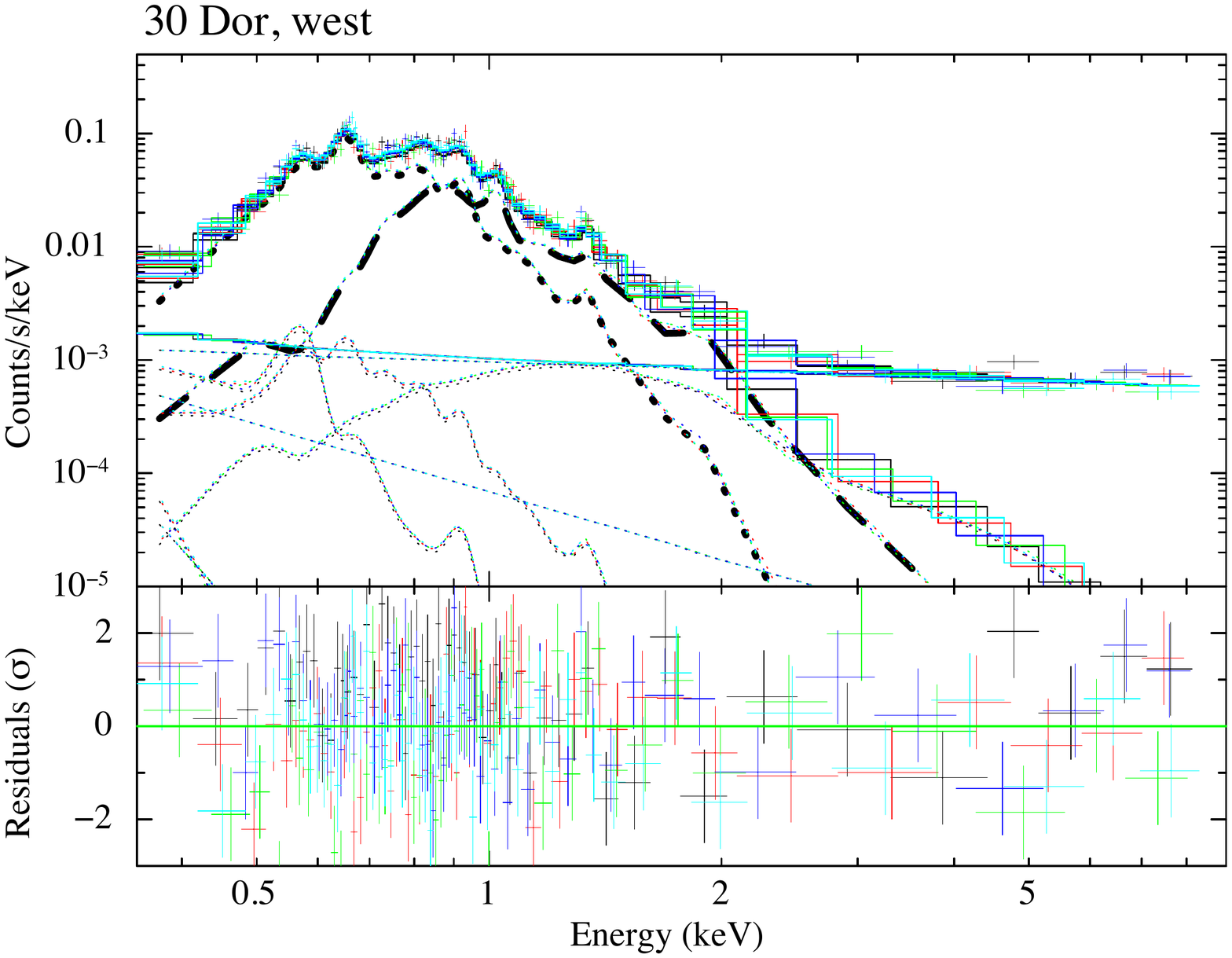}
\includegraphics[width=.5\textwidth,trim=80 40 110 30,clip]{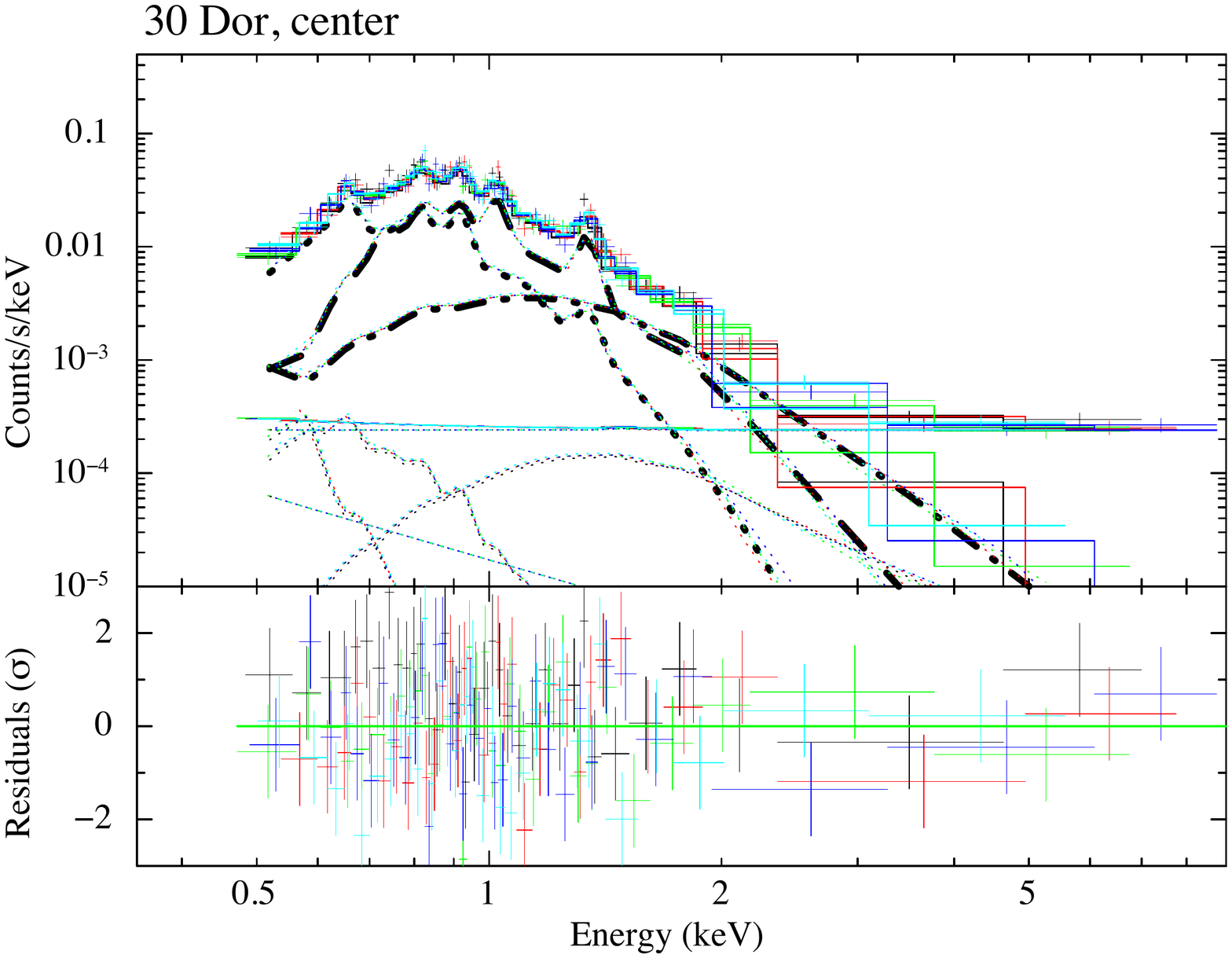}\\
\includegraphics[width=.5\textwidth,trim=80 40 110 30,clip]{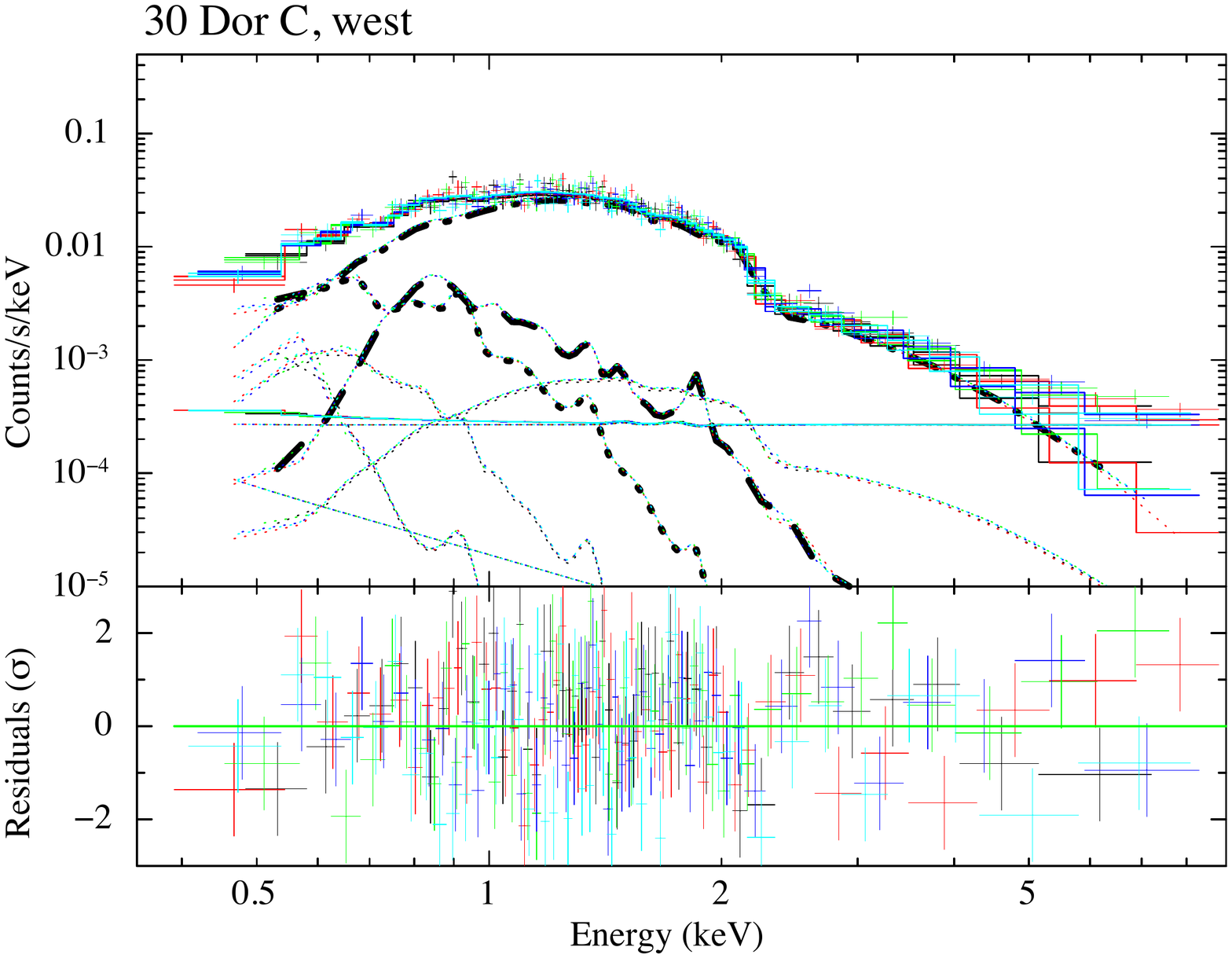}
\includegraphics[width=.5\textwidth,trim=80 40 110 30,clip]{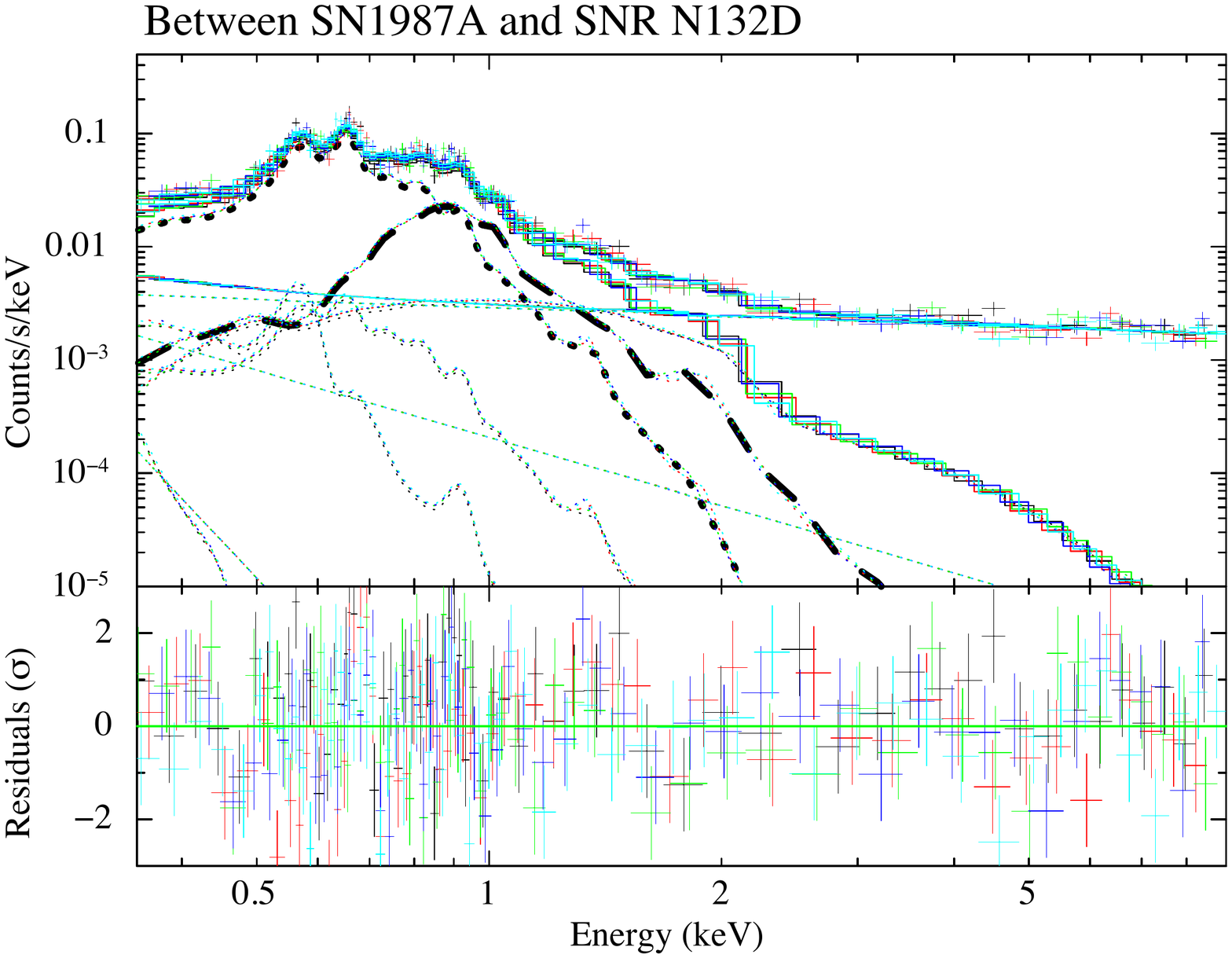}
\caption{\label{spectra}
Spectra of four regions: in the western shell inside 30~Dor (upper left),
close to the centre of 30~Dor (upper right),
in the western part of the non-thermal shell of 30~Dor~C (lower left), and
a region between SN 1987A and SNR N132D with no bright structure in
\ha\ emission (lower right). Spectra of all TMs are shown in different colours.
Thin dashed lines indicate the spectral components 
of the astrophysical background spectrum.
The straight line shows the particle background.
The source components
are highlighted with thick lines (dotted: lower-temperature \vapec1, 
dashed: higher-temperature \vapec2, dash-dotted: \pow).
}
\end{figure*}

\begin{table*}
\centering
\caption{\label{spectab}
Fit parameters. The ranges in brackets are 90\% confidence intervals.
}
\begin{tabular}{llllll}
\hline
Model     & Parameter                 & 30~Dor west           & 30~Dor centre       & 30~Dor~C west       & Between SNRs           \\
\hline
\tbvarabs & \nh$_\mathrm{\,LMC}$ [10$^{22}$ cm$^{-2}$] & 0.32   (0.28 -- 0.37) & 0.55 (0.48 -- 0.61) & 0.54 (0.49 -- 0.61) & 0.083 (0.074 -- 0.096) \\
\vapec1   & $kT_1$ [keV]              & 0.21   (0.19 -- 0.22) & 0.20 (0.18 -- 0.22) & 0.17 (0.14 -- 0.22) & 0.22  (0.21  -- 0.23)  \\
\vapec1   & $norm_1$ [$10^{-3}$ cm$^{-5}$]      & 2.5    (1.9  -- 3.6)  & 3.0  (1.7  -- 5.1)  & 0.78 (0.29 -- 2.0)  & 0.62  (0.57  -- 0.69)  \\
\vapec2   & $kT_2$ [keV]              & 0.73   (0.67 -- 0.78) & 0.40 (0.36 -- 0.46) & 0.62 (0.28 -- 1.1)  & 0.81  (0.75  -- 0.85)  \\
\vapec2   & Ne (solar)                & 2.8    (2.1  -- 3.5)  & 1.7  (1.2  -- 2.7)  & 0.5 (N/A)           & 1.7   (0.5   -- 2.8)   \\
\vapec2   & Mg (solar)                & 1.0    (0.78 -- 1.3)  & 1.3  (0.94 -- 1.9)  & 0.5 (N/A)           & 0.5 (N/A)              \\
\vapec2   & $norm_2$ [$10^{-4}$ cm$^{-5}$]      & 2.2    (2.0  -- 2.5)  & 5.0  (3.0  -- 7.5)  & 0.53 (0.13 -- 2.0)  & 0.83  (0.73  -- 0.94)  \\
\pow      & photon index                     & not constrained                & $>$2.5  & 2.4 (2.3 -- 2.6)  & not constrained  \\
\pow      & $norm$ [$10^{-5}$ photons & $<$3.5                & 8.7  (6.7  -- 11.)  & 32.  (31.  -- 34.)  & $<$1.6  \\
& keV$^{-1}$ cm$^{-2}$ s$^{-1}$] \\
& $\chi^2$  & 373 & 197 & 322 & 403 \\
& d.o.f.    & 295 & 175 & 267 & 328 \\
\hline
\end{tabular}
\end{table*}

We analysed the spectra using \texttt{XSPEC} version 12.11.1.
We modelled the spectrum of the diffuse emission with a combination
of two thermal plasma models and a power-law, absorbed by
gas in the Milky Way and in the LMC. Studies of the hot ISM
in the Milky Way, the Magellanic Clouds, and the nearby galaxies
with \chandra, \suzaku, or \xmm\ 
\citep[e.g.,][and references therein]{2010ApJS..188...46K,2020A&A...637A..12K} 
have shown that the diffuse X-ray emission in the
ISM of normal galaxies requires at least two thermal plasma components with
different temperature: $kT_1 \approx$ 0.2 keV consistently in most cases, 
most likely 
emission from hot ISM in equilibrium and from unresolved stellar sources,
and $kT_2 > 0.5$ keV from regions, which experienced recent heating, i.e.,
\hii\ regions, superbubbles, and SNRs, or also include unresolved
binaries. 
For the plasma emission we tried both the collisional ionisation equilibrium model \vapec\footnote{\url{http://atomdb.org/}} and non-equilibrium ionisation model \vnei\ \citep[][and references therein]{2001ApJ...548..820B}. 
Since the lower-temperature component has an ionisation timescale 
$\tau = n_e t$ of $\sim10^{13}$ s cm$^{-3}$ and is thus consistent with 
collisional ionisation equilibrium, while that of the higher-temperature
component is not well constrained, we  decided to use the results 
obtained with two \vapec\ models for further discussion.
In addition, some SNRs and a few superbubbles are also known  to emit 
significant non-thermal X-ray emission, with 30~Doradus C (30~Dor~C)
located close to SN 1987A being one of the only two non-thermal 
superbubbles known so far in the Local Group of galaxies. 
We have therefore included a power-law component to model the emission in regions like 30 Dor C and to verify if non-thermal emission can also be detected outside the known non-thermal sources.
For the foreground absorption, we have included two components, one for
the absorbing column in the Milky Way \nh$_\mathrm{, Gal}$, which is fixed
to the value from the newly calculated map of the Galactic $N_{\mathrm{H, cold, Gal}}$ 
(see Sect.\,\ref{nhcold}), and another component for the LMC with 
0.5 $\times$ solar
abundances.
For the \nh, we use the model \tbvarabs\ \citep{2000ApJ...542..914W}.

For the fits we first set all element abundances to the average value of
0.5 $\times$ the solar values
\citep{1992ApJ...384..508R}. 
We assumed the element abundances reported by \citet{1989GeCoA..53..197A}.
In the next step we freed the abundances of O, Ne, and Mg in
the hot thermal plasma emission component one
by one since there is significant emission from thermal plasma up to 
2.0 keV.
We verified if the fit improves with the new free parameter using the F-test
statistics. If the change was not significant, the parameter was set back
to 0.5 $\times$ solar. 
Figure \ref{redchi} shows the distribution of the values for $\chi^2$/d.o.f.

In Fig.\,\ref{spectra} we show four example spectra with the best-fit models. 
The upper left spectrum is 
taken in the western wing of the giant \hii\ region 30~Dor, while the upper right panel shows the spectrum of
the central region of 30~Dor.
The lower left spectrum was extracted in the western part of the
non-thermal shell of 30~Dor~C, the superbubble located south-west of
30~Dor. In addition, we also show one of the spectra taken in a less
active region between SN 1987A and SNR N132D, in which no bright
\ha\ structures are observed. The best-fit parameters for the spectra are 
listed in Table \ref{spectab}.


\begin{figure*}
\centering
\includegraphics[width=.495\textwidth]{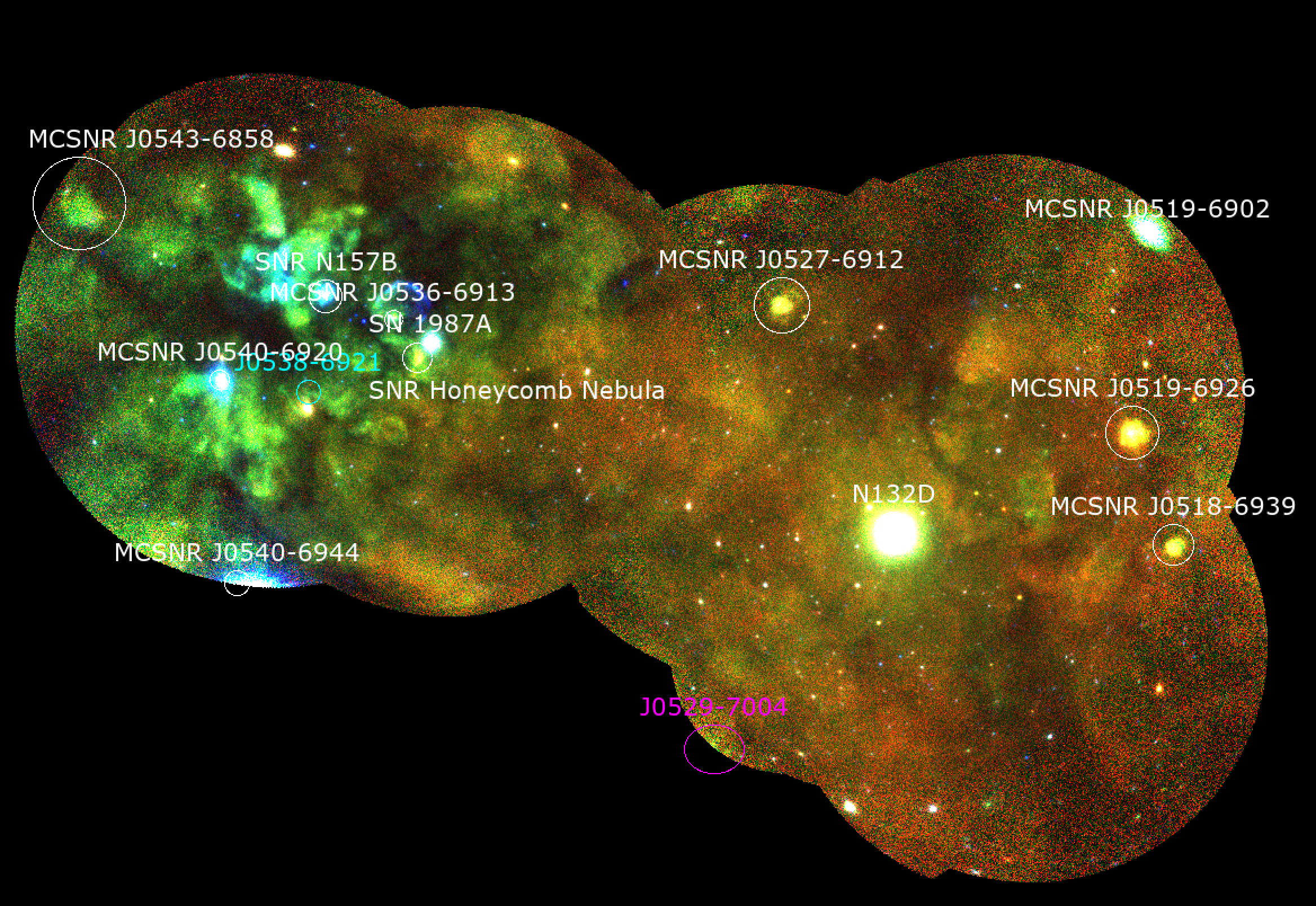}
\includegraphics[width=.495\textwidth]{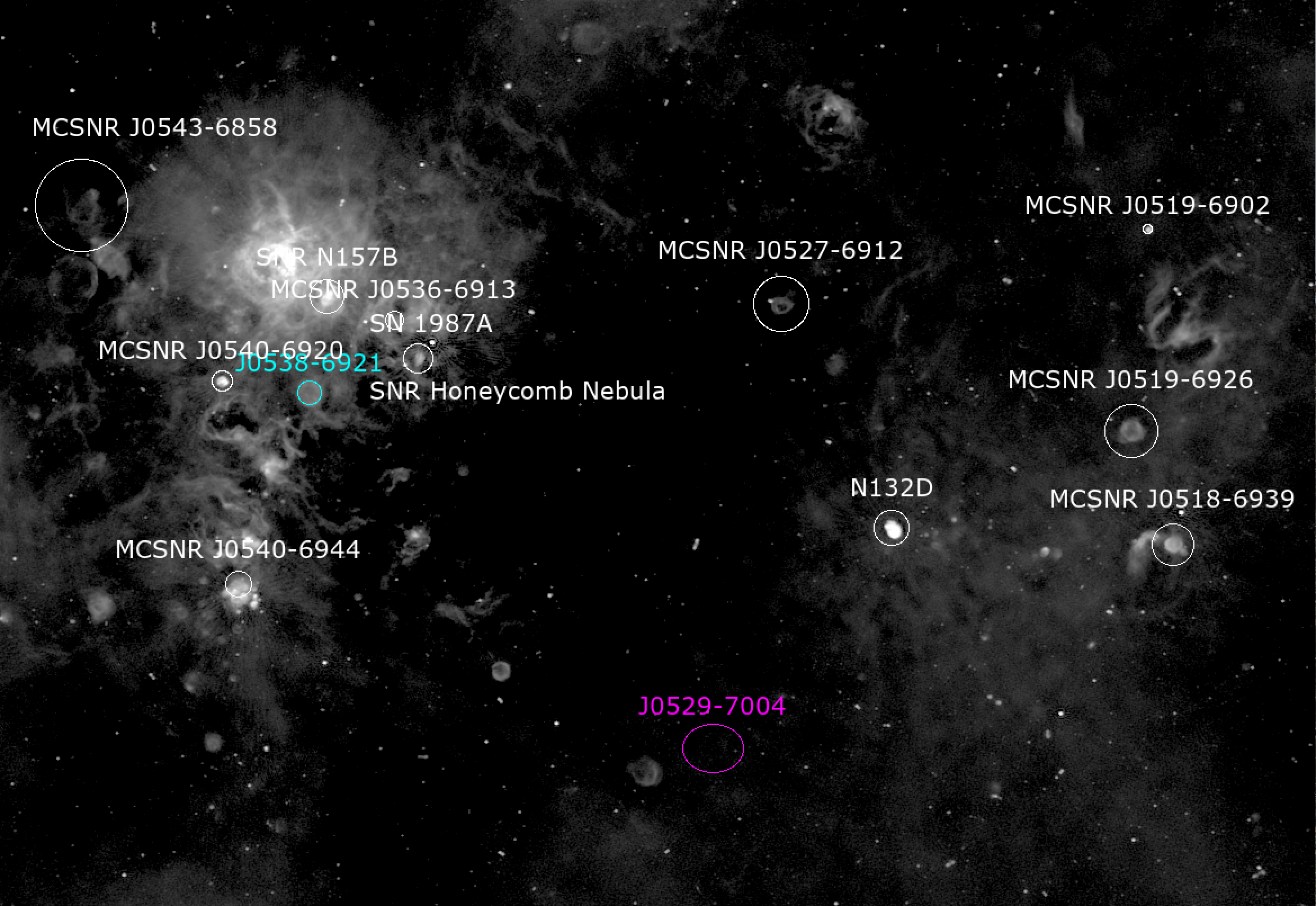}
\caption{\label{xrsnrs}
eROSITA mosaic images in three
colours (same as in Fig.\,\ref{mosaic}, left) and ASKAP 888 MHz image (Pennock et al. 2021, accepted). Known SNRs are marked in white, while the SNR candidates J0529--7004 and J0538--6921 are shown in magenta and cyan, respectively.}
\end{figure*}

\begin{figure}
\centering
\includegraphics[width=.49\textwidth]{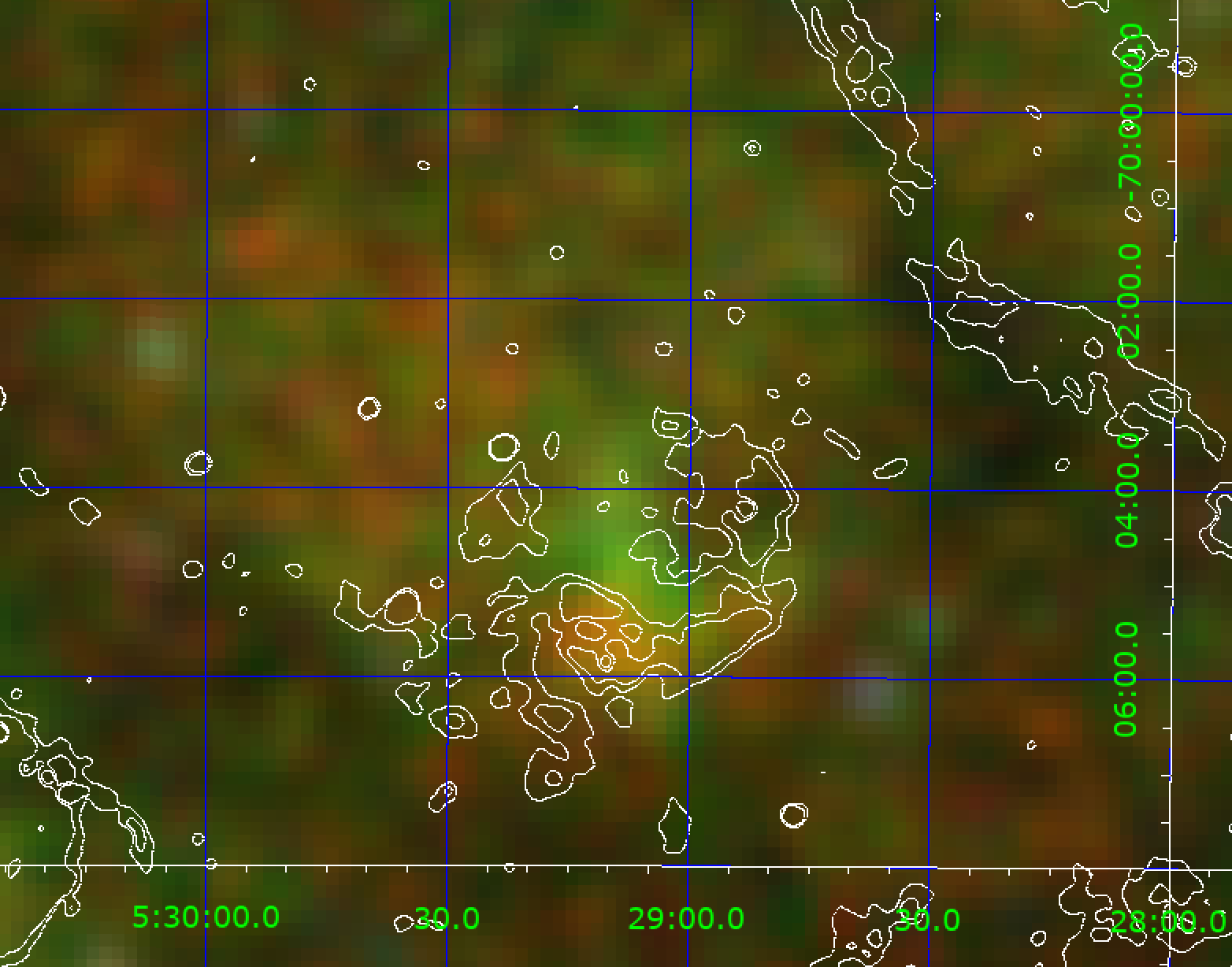}
\caption{\label{ima:snrj0529}
Three-colour eROSITA image of SNR J0529--7004 (red: 0.2 -- 0.5 keV, green = 0.5 -- 1.0 keV, blue = 1.0 -- 2.0 keV)
with contours of \sii\ emission (MCELS).
}
\end{figure}

\begin{figure}
\centering
\includegraphics[width=.49\textwidth,trim=80 45 110 35,clip]{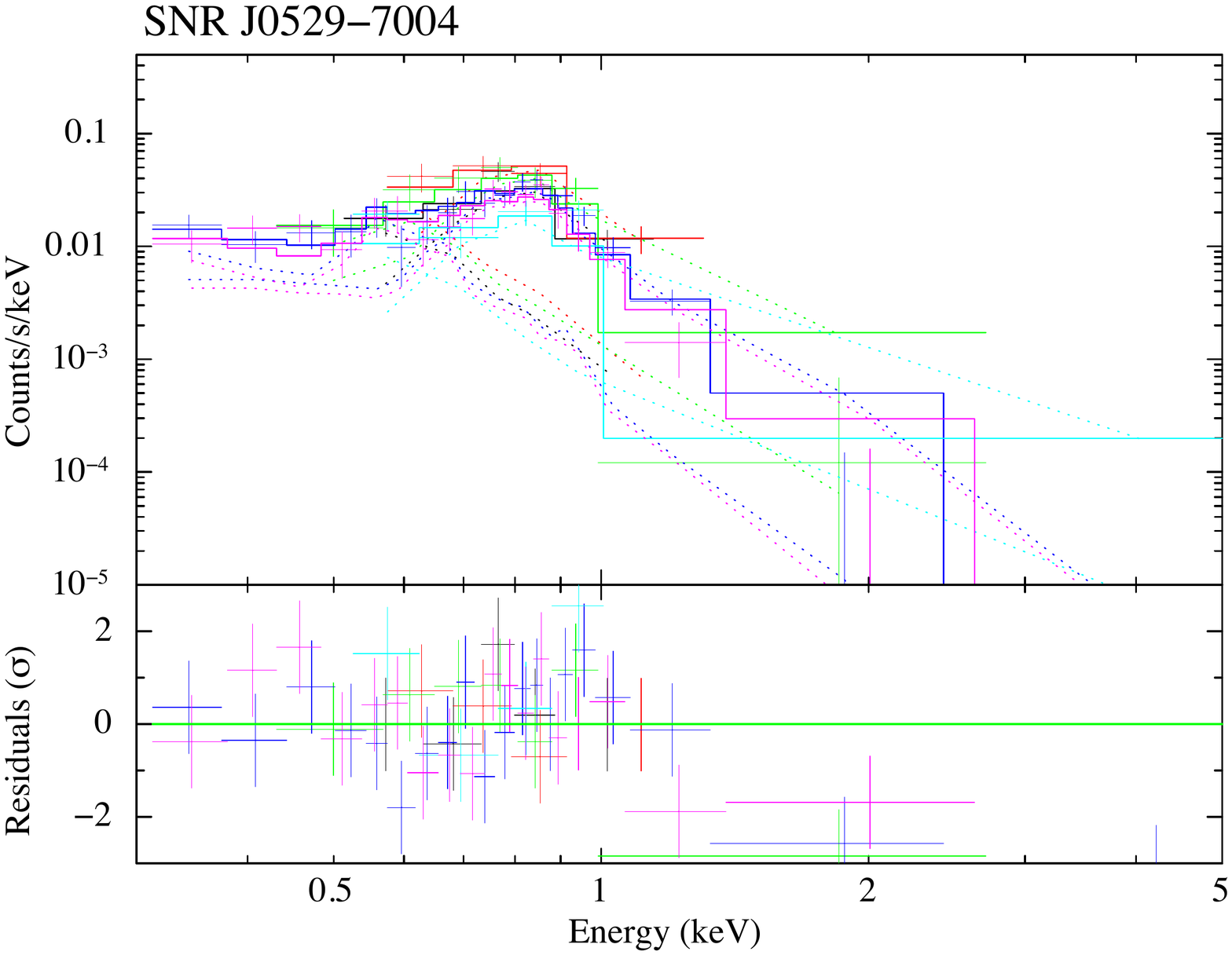}
\includegraphics[width=.49\textwidth,trim=80 45 110 35,clip]{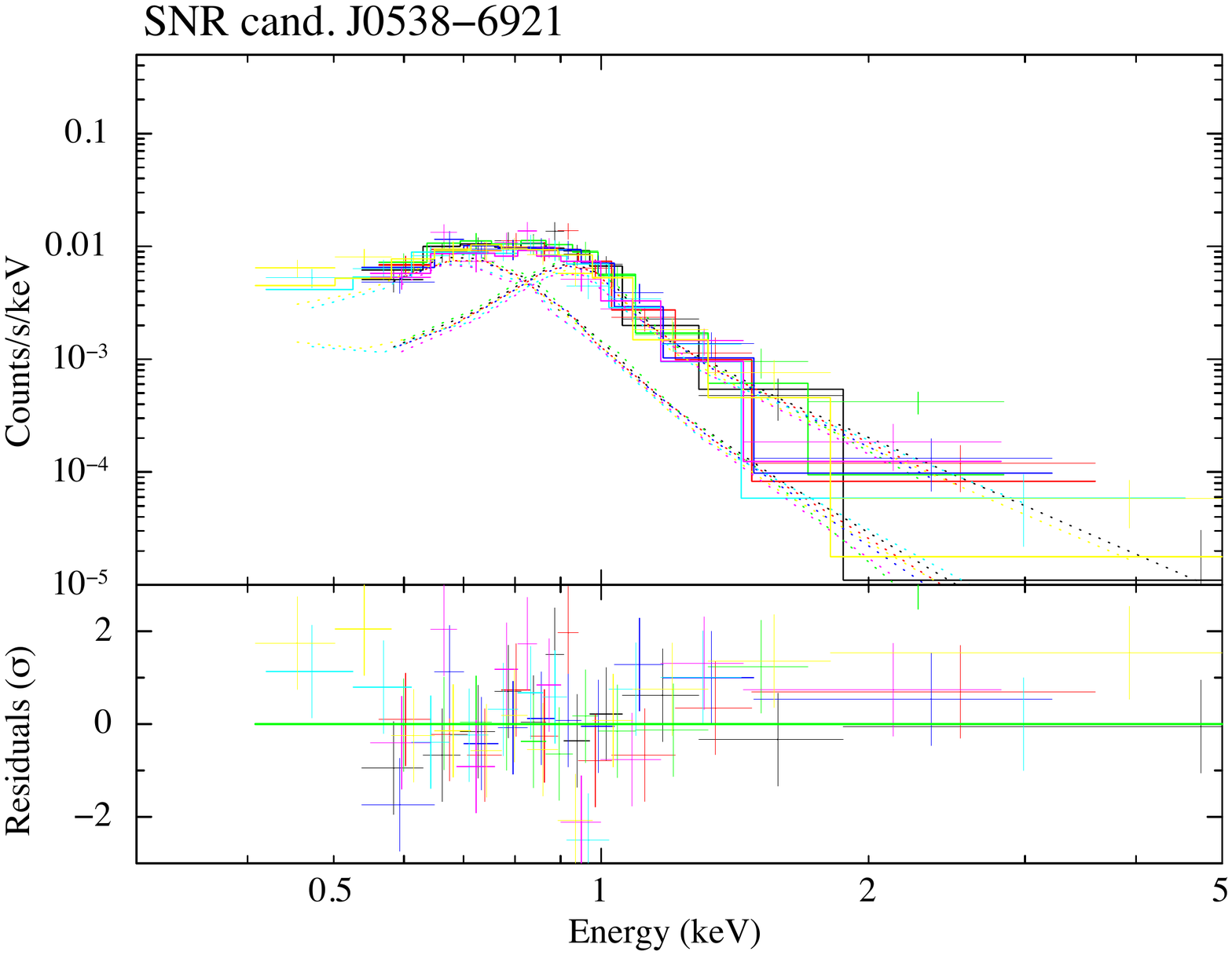}
\caption{\label{spec:snrs}
eROSITA spectra of SNR J0529--7004 (upper panel, ObsID 700185) and SNR candidate J0538--6921 (lower panel, ObsID 700161) with the best-fit 2 \vapec\ models.}
\end{figure}

\section{Massive Stars in 30 Doradus}

The 30 Doradus region is known to host a large number of very massive stars \citep{1960MNRAS.121..337F}, including the compact cluster of young massive stars RMC 136 in the centre. It contains the Wolf-Rayet star R136a1, which is one of the most massive stars detected so far \citep[][]{2010MNRAS.408..731C}. Compact clusters of young massive stars 
are expected to be very efficient cosmic ray accelerators \citep[see for review][]{2014A&ARv..22...77B,2020SSRv..216...42B}.  
In X-rays bright diffuse emission has been detected from the central region in addition to point sources, which were resolved with \chandra\
\citep{2006AJ....131.2140T,2006AJ....131.2164T}

With eROSITA, two bright sources are detected in the central region, one at the position of the star cluster RMC 136 and another source at the position of the Wolf-Rayet stars RMC 139 and RMC 140. We have extracted the spectra of these two extended sources. The emission in both regions are well reproduced with  thermal emission from a non-equilibrium ionisation plasma (\vnei) and non-thermal emission (\pow). The temperature of the plasma is $kT >$ 1 keV and thus higher than in the surrounding regions with diffuse X-ray emission. The photon index of the power-law component is low $\Gamma = 1.3$ indicating a hard X-ray spectrum (Table \ref{rmcspectab}). Using the eROSITA spectrum we obtain a significantly enhanced abundance for Mg in both regions and for Si for RMC 136. Other elements are not well constrained due to poor photon statistics for the emission lines.

\begin{table}
\centering
\caption{\label{rmcspectab}
Fit parameters for the emission at RMC 136 and RMC 139/140. The ranges in brackets are 90\% confidence intervals.
}
\begin{tabular}{llll}
\hline
Model & Parameter & RMC 136 & RMC 139/140 \\
\hline
\vnei & $kT$ [keV] & 1.9 (1.4 -- 2.9) & 1.3 (1.1 -- 1.7) \\
\vnei & $\tau$ [10$^{11}$ s cm$^{-3}$] & 0.9 (0.6  -- 1.5)  & 1.4 (0.9 -- 2.3) \\
\vnei & Mg (solar)  & 3.9  (2.7  -- 5.6)  & 3.5  (2.5  -- 4.6) \\
\vnei & Si (solar)  & 3.6 (2.1 -- 5.4)  & 0.5  (N/A) \\
\pow  & photon index & 1.3 (1.1 -- 2.4)  & 1.3 (1.2 -- 1.5) \\
& $\chi^2$ & 164 & 200 \\
& d.o.f. & 97 & 104 \\
\hline
\end{tabular}
\end{table}

\section{Supernova Remnants}

In the part of the LMC, which has been observed with eROSITA in the early phase, there are several known SNRs, which can all be confirmed in the eROSITA data (Fig.\,\ref{xrsnrs}). In addition, there are two sources which are known to be SNR candidates (one radio and one optical source). The analysis of the eROSITA emission of these objects are presented in the following.

\subsection{SNR candidate J0529--7004}

SNR candidate J0529--7004 is an arc-like structure seen in optical emission-line images with a high flux ratio of \sii/\ha\ = 1. Therefore, it has been suggested to be an SNR candidate by \citet{2021MNRAS.500.2336Y}. An X-ray source was detected with ROSAT at its position \citep{1999A&AS..139..277H}. With eROSITA, the position was observed in observation 700185.
There is faint extended X-ray emission as can be seen in the three-colour image in Fig.\,\ref{ima:snrj0529}. 
We extracted spectra at the position of the optical SNR candidate. The local background was taken from a nearby region with no significantly enhanced emission
and subtracted from the source spectrum.
The eROSITA spectra are shown in Fig.\,\ref{spec:snrs} (upper panel). 
First, we fit the spectra with one thermal plasma model \vapec\ or \vnei, which did not yield a good fit.
We, therefore, fit the spectrum again with a two \vapec\ model and obtain 
the best-fit with $\chi^2 = 132$ and 61 degrees of freedom (d.o.f) with the following parameters:
$kT_1$ = 0.20 (0.13 -- 0.27) keV, $kT_2$ = 0.68 (0.62 -- 0.74) keV. The foreground column density tends to be zero with an upper limit of \nh\ = $1.6 \times 10^{20}$ cm$^{-2}$. The best-fit model yields a flux (absorbed) of $F_\mathrm{X} (0.2 - 10.0 \mathrm{~keV}) = (1.4\pm1.0) \times 10^{-13}$ erg s$^{-1}$ cm$^{-2}$.
Due to poor photon statistics, element abundances could not be determined and were all set to 0.5 times solar. 
The X-ray emission and the optical \sii/\ha\ flux ratio confirm this source to be an SNR.

\subsection{SNR candidate J0538--6921}

SNR candidate J0538--6921 was detected in radio and classified as an SNR candidate \citep[][and references therein]{2017ApJS..230....2B} with a radio spectral index of $\alpha = -0.59 \pm 0.04$. Very faint filaments can be seen in the optical, but a clear structure indicative of an SNR is missing. In the eROSITA data (ObsID 700161), there is diffuse X-ray emission at the position of the radio candidate, but without a clear structure that could be identified as an SNR.
We extracted the spectra at the position of the radio source and in a nearby background region (Fig.\,\ref{spec:snrs}, lower panel). However, since there is an X-ray bright foreground star located in the south close to the object, it was not possible to extract the X-ray spectrum at the entire position of the radio source. 
The background spectrum was  subtracted from the source spectrum.
The best-fit parameter values for a two \vapec\ model 
with $\chi^2 = 92$ and 77 degrees of freedom (d.o.f)
are: $kT_1$ = 0.30 (0.27 -- 0.33) keV, $kT_2$ = 0.95 (0.89 -- 1.1) keV. Also in this case the foreground column density is not well constrained. The X-ray emission is very faint with a flux (absorbed) of $F_\mathrm{X} (0.2 - 10.0 \mathrm{~keV}) = (6\pm2) \times 10^{-14}$ erg s$^{-1}$ cm$^{-2}$.
Due to the lack of a clear structure in X-rays and the very faint flux, this object cannot be confirmed as an SNR.
Since this source is located in a region with enhanced diffuse emission both in the images of the optical line emission (see Fig.\,\ref{imanorm}, lower right) and those in X-rays, it is difficult to identify a possible SNR. In addition, the diffuse emission in the optical and X-rays suggest that the region is filled with ionised gas and has most likely been heated by interstellar shocks in the past. In such a region, the SNR would be expanding in a medium with a density that is lower than in an unshocked medium, which would explain the lack of obvious X-ray and optical emission from the SNR.

\section{Discussion}

\begin{figure*}
\centering
\includegraphics[width=\textwidth]{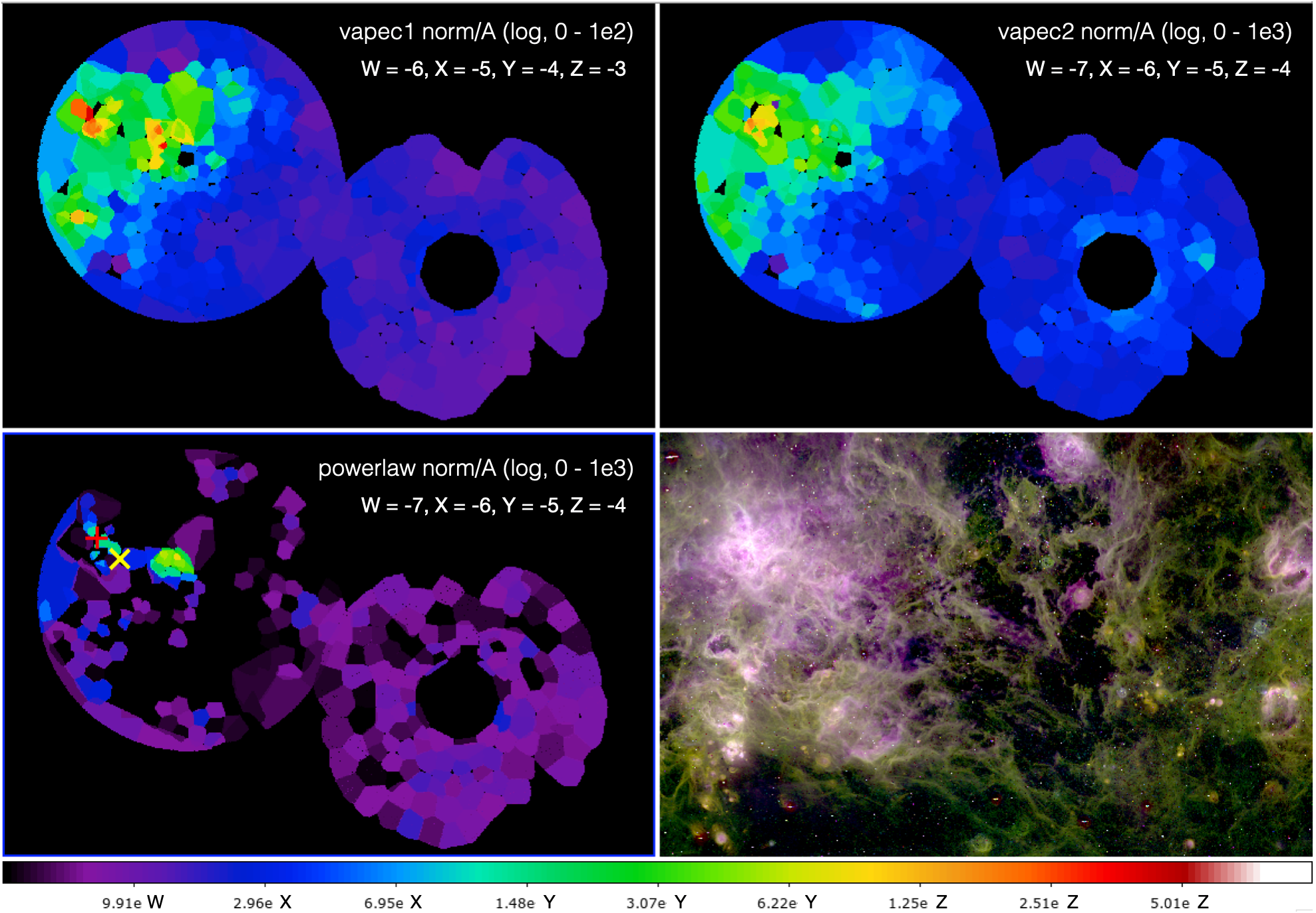}
\caption{\label{imanorm}
Parameter maps of the normalisation per arcmin$^2$, illustrating the
surface brightness of the diffuse emission,
for the two thermal components (\vapec1, \vapec2)
and the \pow\ component. The images are shown in log-scale.
The letters W, X, Y, Z in the labels correspond to the exponents of the marks of the color scale.
The red plus sign indicates the position of the star cluster RMC 136,
the yellow cross that of SNR/PWN N157B.
The lower right panel shows \ha\ (red), \sii\ (green), and
\oiii\ (blue) images of MCELS in three-colour presentation.
}
\end{figure*}

We have created maps for all fit parameters by filling the extraction regions
with the parameter values of the best-fit models. The parameter maps for the 
normalisation per arcmin$^2$ for the three source emission componets
are shown in Fig.\,\ref{imanorm}.
Normalisation in XSPEC for the thermal model \vapec\ is defined as
\begin{equation}\label{eq:norm}
norm_{\vapec} = \frac{1}{10^{14} \times (4\pi D_{\rm LMC}^2)} \int n_e n_{\mathrm H} dV \quad [\mathrm{cm}^{-5}]
\end{equation}
with $n_e$ and $n_{\mathrm H}$ being the electron and hydrogen densities, respectively.
For the powerlaw model the normalisation is
$norm_{\pow}$ = number of photons keV$^{-1}$ cm$^{-2}$ s$^{-1}$ at 1 keV.
For the images in Fig.\,\ref{imanorm} the normalisation of the fit has been
divided by the size of the extraction region $A$ in arcmin$^2$.

\subsection{Thermal Component}

In all regions, there is significant emission from thermal plasma.
We calculated the average temperature and its standard deviation from the parameter maps for $kT_1$ and $kT_2$, in which each pixel is filled with the best-fit parameter values. We thus obtain
a mean temperature of $kT_1$ = 0.22 keV with a standard deviation of 
$\sigma_{kT1}$ = 0.02 keV for the lower-temperature component and $kT_2$ = 0.74 keV and $\sigma_{kT2}$ = 0.10 keV for the higher-temperature component. 
These temperature values are consistent with the results of the analysis of \xmm\ observations of southeastern parts of the LMC (regions around 30~Dor, in the X-ray spur, and west of them) by 
\citet{2021A&A...648A..90K}.
As can be seen in  Fig.\,\ref{imanorm} (upper panels),
the normalisation per area
of the lower-temperature component (\vapec1) is 
one order of magnitude higher
than that of the 
higher-temperature component (\vapec2).

From magnitude and colour measurements of stars in the LMC, the thickness of the disk was determined to be $d$ = 3$\pm$1 kpc at the position of SNR N132D by \citet{2009A&A...496..399S} and \citet{2012A&A...537A.106R}. The 
volume $V$ of the plasma in each of the spectral analysis regions can be written as
\begin{equation}
V \approx \frac{A}{60^2} \frac{\pi^2}{180^2}~D_{\rm LMC}^2~d
\end{equation}
where $A$ is the area of the extraction region in arcmin$^2$, $D_{\rm LMC}$ = 50 kpc is the distance to the LMC, and $d$ = 3 kpc is the thickness of the disk
By assuming a homogeneous depth of the volume, introducing a filling factor $f$ for the plasma, and using the relation $n_e = 1.2 n_{\mathrm H}$, we can write 
\begin{eqnarray}
\label{eq:density}
n_{\mathrm H} &\approx& \left(\frac{1.2 \times 10^{14} \times (4\pi D_{\rm LMC}^2)~norm_{\vapec}}{f V}\right)^{\frac{1}{2}} \nonumber \\
&=& \left(\frac{5.6 \times 10^{22} \times norm_{\vapec}/A}{\pi~f~d }\right)^{\frac{1}{2}}. 
\end{eqnarray}
With the mean value of $norm_{\vapec1}/A = (1.6\pm1.0) \times 10^{-5}$ cm$^{-5}$ arcmin$^{-2}$ (fit uncertainty of $\sim$10$^{-4}$ and standard deviation of $0.4 \times 10^{-5}$)
for the brighter, lower-temperature \vapec1 component in the western regions around SNR N132D, we
get a mean density of
%
$n_{\mathrm H, 1} = (3.1\pm1.5) \times 10^{-4} f^{-1/2}$ cm$^{-2}$.
We assume that the emitting plasma can be regarded as an ideal gas with the pressure
\begin{equation}
P/k_\mathrm{B} = (n_e + 1.1 n_\mathrm{H}) T
= 2.3 n_\mathrm{H} T
\end{equation}
where $k_\mathrm{B}$ is the Boltzman constant.
The temperature $T_1$ of  \vapec1 and the hydrogen number density $n_\mathrm{H,1}$ derived from the normalisation  yields a mean pressure of
$P_1/k_\mathrm{B}  = (1.9\pm1.1) \times 10^3 f^{\frac{1}{2}} \mathrm{cm}^{-3} \mathrm{K}$ in the western regions. This pressure is consistent with the thermal pressure in the disk of the Milky Way \citep{2005ARA&A..43..337C}.

The higher normalisations of the thermal emission models in the eastern regions around 30~Dor compared to the regions around SNR N132D imply that the emission is 
brighter in and around 30~Dor as well as in the southeast than towards the west.
Those are the regions, which also show bright line emission in the optical,
as can be confirmed in the images of the Magellanic Clouds Emission Line Survey
\citep[MCELS, see Fig.\,\ref{imanorm}, lower right,][]{2004AAS...20510108S}.
In the eROSITA mosaic images (Fig.\,\ref{mosaic}), the emission in these 
regions appear to be fainter than or as faint as in the west. 
At the same time, the regions also appear 
green in the three-colour presentation, suggesting that the observed variation
in surface brighness and colour is due to higher absorption. This will be 
further discussed in Sect.\,\ref{absorption}.

\subsection{Non-thermal Component}

Non-thermal emission is observed significantly in the superbubble 
30~Dor~C (see Fig.\,\ref{imanorm}, lower left, and Fig.\,\ref{spectra}, lower left). The photon index obtained from the eROSITA data is $\Gamma$ = 2.5$\pm$0.5, which is 
in agreement with former studies 
\citep[e.g.,][and references therein]{2015A&A...573A..73K,2019A&A...621A.138K}.
Recently, 
in a combined XMM-Newton and NuSTAR analysis, \citet{2020ApJ...893..144L} showed that the complete shell of 30 Dor C is detected up to 20 keV.
The authors found that contrary to prior studies, their Region D 
(part of
30 Dor C West here) requires a thermal component of $kT = 0.86 \pm 0.01$ keV which is  consistent with the eROSITA fit with the temperature of  0.62 (0.28 -- 1.1) keV given in Table \ref{spectab}. The photon indices of a power law component obtained by eROSITA 
for the entire 30 Dor C West region
is $\Gamma$ = 2.4 (2.3 -- 2.6), which is somewhat higher than the value of $2.12 \pm 0.02$ obtained for region D, but consistent with $2.39 \pm 0.03$ obtained for region C, which were both covered by our 30 Dor C West region.
The SNR MCSNR J0536--6913 which was found earlier with \xmm\ \citep{2015A&A...573A..73K,2018ApJ...864...12B} and is apparent in the eROSITA image (Figure \ref{xrsnrs}, left) was not detected above 8 keV in the NuSTAR observation.

Synchrotron radiation of very high-energy electrons and positrons is a likely source of a substantial amount of the observed non-thermal X-rays. The high energy leptonic component is accelerated in supernova remnants, superbubbles, and pulsar wind nebulae and can be produced by inelastic collisions of cosmic ray nuclei. The high-energy leptons producing the synchrotron X-rays simultaneously up-scatter photons from ambient radiation field. The inverse-Compton radiation    
from multi-TeV leptons producing the observed X-ray synchrotron apparently contribute to the TeV gamma-ray emission 
detected from 
the pulsar wind nebula (PWN) N157B and the superbubble 30 Dor C
by the ground-based Cherenkov gamma-ray telescope H.E.S.S. \citep[see, e.g.,][]{2015Sci...347..406H}. 
H.E.S.S.
measured a luminosity of $\sim 0.9\times 10^{35}$ erg s$^{-1}$ for 30 Dor C
for the 1 -- 10 TeV gamma-rays with a power-law distribution of a photon index $2.6 \pm 0.2$, which is consistent with the photon index of X-rays detected by eROSITA in 30 Dor C west (see Table \ref{spectab}). The observed TeV emission apart from the inverse-Compton mechanism mentioned earlier could originate from the decay of neutral pions produced in the inelastic collisions of relativistic nuclei with the ambient matter (so-called hadronic scenario). While the expected fluxes in the leptonic scenario can explain the observed fluxes of X-rays and TeV gamma-rays as well as the measured widths of the non-thermal X-ray filaments \citep{2019A&A...621A.138K}, one can still not exclude the hadronic origin of the observed gamma-rays  \citep[see, e.g.,][]{2015Sci...347..406H}.

There is also significant non-thermal emission in the central region of 30~Dor
(see Fig.\,\ref{spectra}, upper right, Table \ref{spectab}). This non-thermal
component can indicate some contamination by emission from the 
composite SNR with a PWN N157B, located southwest of the region
(yellow cross in  Fig.\,\ref{imanorm}, lower left). The emission of N157B itself 
was removed when point and point-like sources
were cut out from the event files.
The X-ray spectrum of N157B consists of thermal emission of the SNR and the dominant non-thermal powerlaw emission from the pulsar and the PWN with a photon index of $\Gamma$ = 2.29 (+0.05,--0.06), while the
pulsar shows a powerlaw emission with $\Gamma$ = 1.73 (+0.11,--0.06)
\citep{2006ApJ...651..237C}. The high lower limit of $\Gamma > 2.5$ for the photon index for the non-thermal emission detected inside 30~Dor suggests that it is not caused by emission from N157B. It should be noted, however, that the gamma-ray spectrum of N157B detected by H.E.S.S. \citep[][]{2012A&A...545L...2H} can be described well by a power-law component with a photon index $\Gamma = 2.8 \pm 0.2_{stat} \pm 0.3_{syst}$ in the energy range between 600 GeV and 12 TeV.

Since non-thermal emission is also found to the north of the
super-star cluster RMC 136 (red plus sign) and possibly in a larger region 
towards east, it might also be caused by particles which were accelerated in 
the shocks of the winds of massive stars inside RMC 136
and are diffusing inside and probably also out of the nebula.
By now diffuse non-thermal X-ray emission was detected with \chandra\ from the most massive Galactic star cluster Westerlund 1 \citep{2006ApJ...650..203M}. 
They estimated the X-ray luminosity to be about $3 \times 10^{34}$ erg s$^{-1}$ and found that while the photon index of the power-law X-ray component is $\Gamma = 2.7 \pm 0.2$  within the circle of 1\arcmin\ radius around the cluster, the spectrum gets harder with $\Gamma = 2.1^{+0.1}_{-0.2}$ in the annulus between 1\arcmin\ and 2\arcmin\ and even $\Gamma = 1.7^{+0.1}_{-0.1}$ between 2\arcmin\ and 3.5\arcmin. Possible mechanisms of the origin of
the non-thermal emission in the clusters of young massive stars were discussed by \citet{2014A&ARv..22...77B}. 

Further studies will be performed in the future when additional data
of the eROSITA All-Sky Survey will be available.

\subsection{Element Abundances}

\begin{figure}
\centering
\includegraphics[width=.49\textwidth]{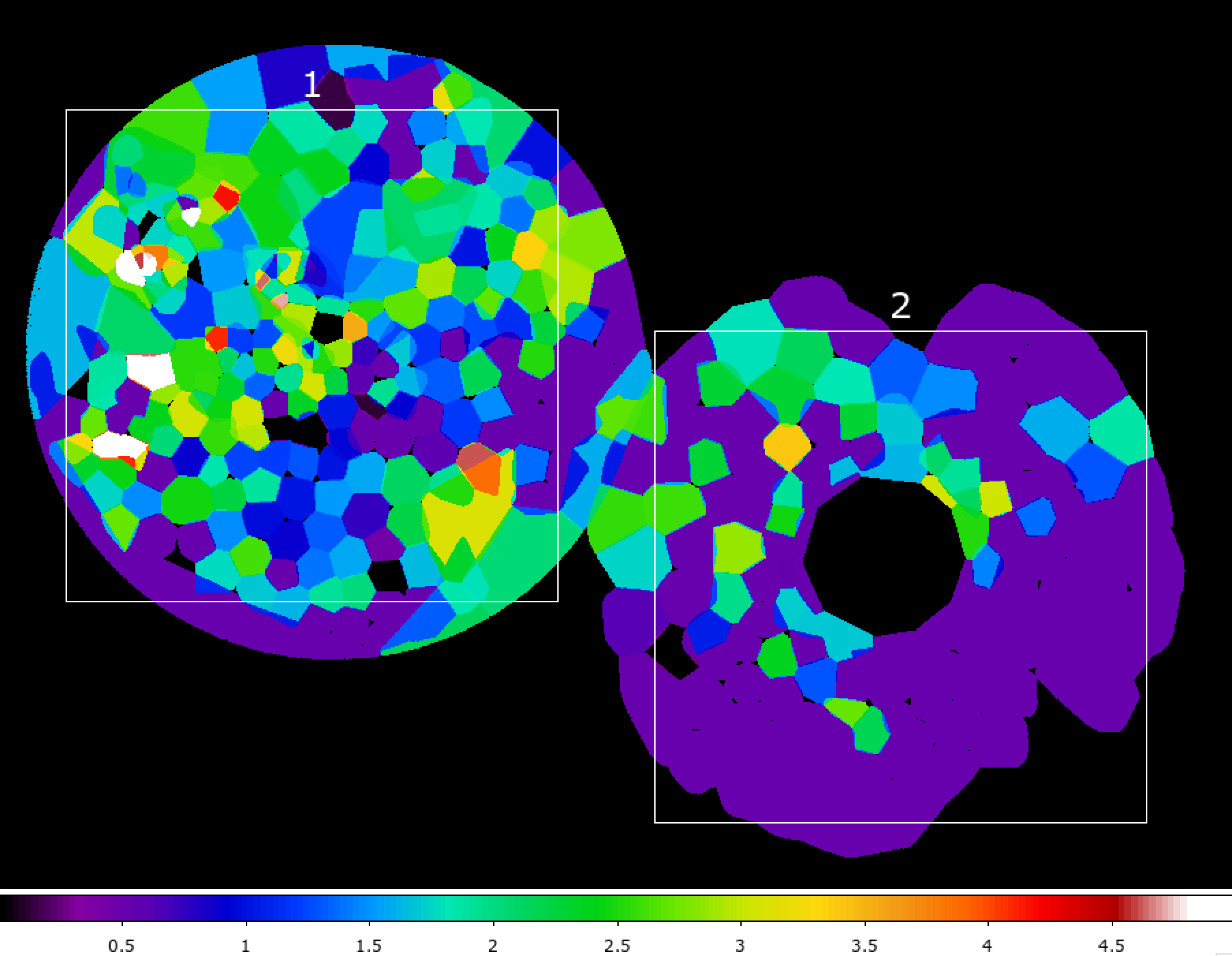}
\caption{\label{imaneabund}
Parameter map of fitted Ne abundances (normalised to solar abundance).
The white boxes are the regions used for the calculation of the star-formation
rate (Fig.\,\ref{sfr}).
}
\end{figure}

\begin{figure}
\centering
\includegraphics[width=.5\textwidth,trim=75 10 80 70,clip]{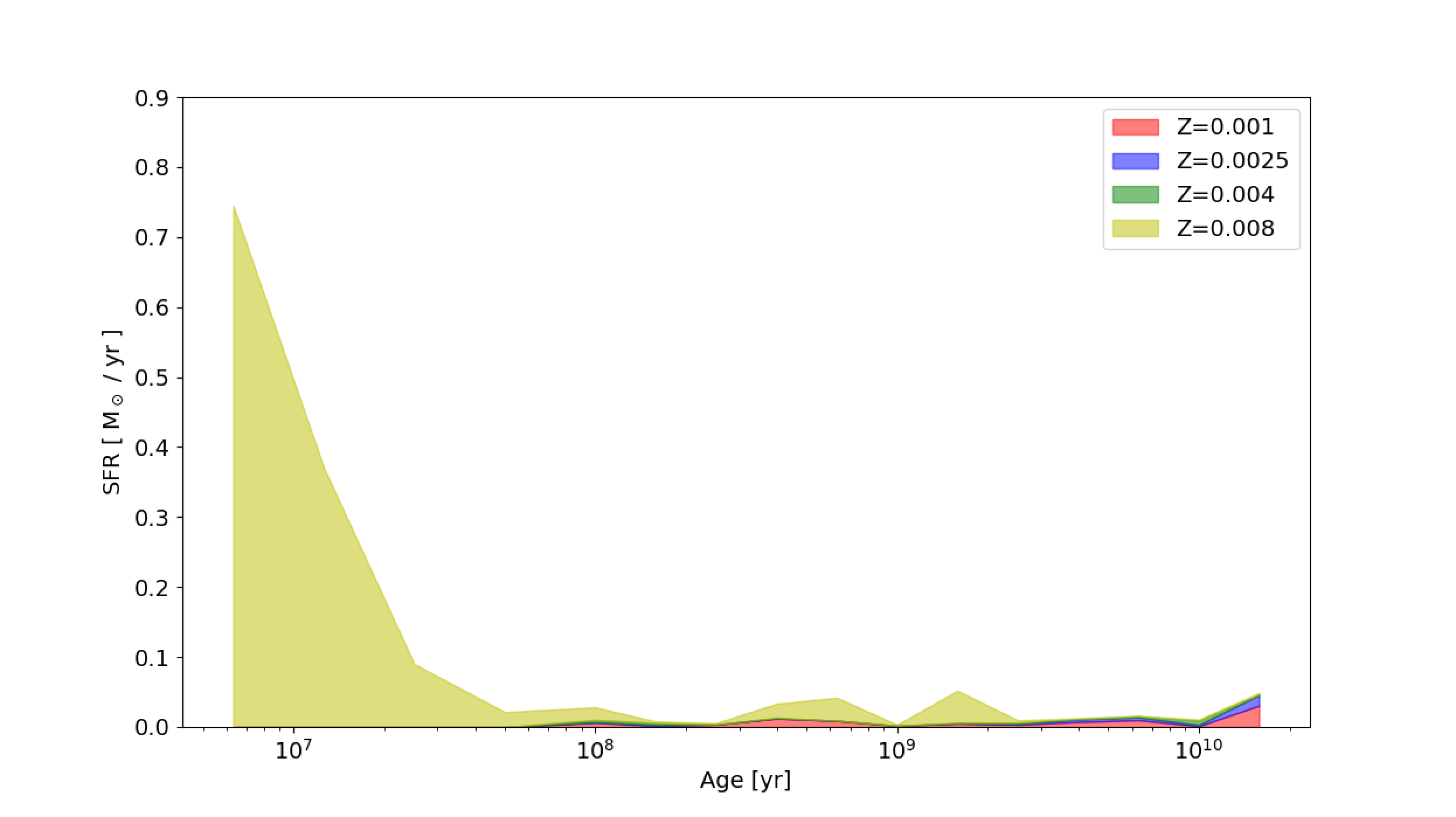}
\includegraphics[width=.5\textwidth,trim=75 10 80 70,clip]{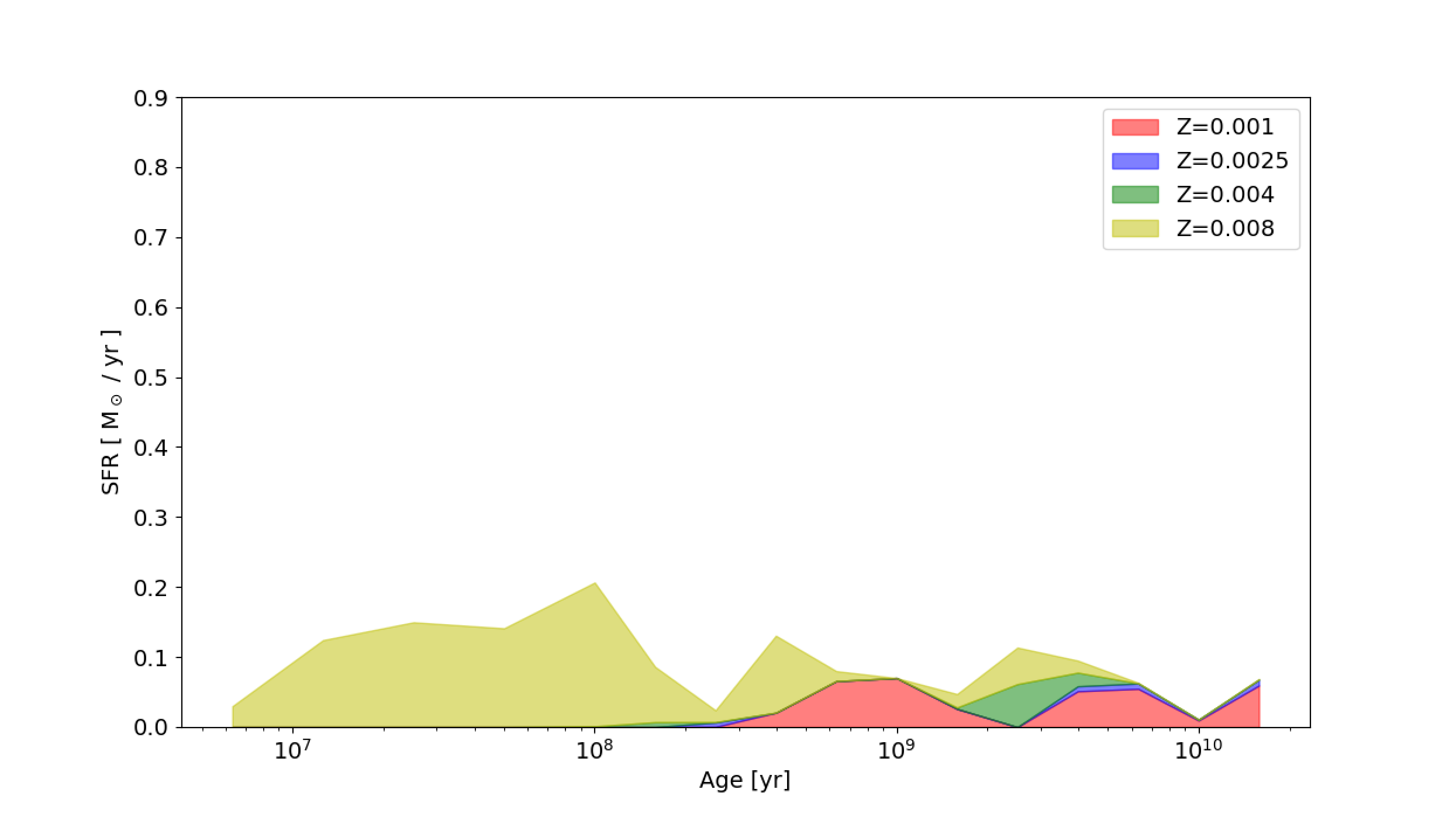}
\caption{\label{sfr}
Star-formation rate from \citet{2009AJ....138.1243H} 
in the regions in Fig.\,\ref{imaneabund} (upper: 1, lower: 2).
}
\end{figure}

The eROSITA spectra suggest that the element abundance of neon is 
enhanced in the regions around 30 Dor (Tab.\,\ref{spectab},
Fig.\,\ref{imaneabund}), while
it is consistent with the typical LMC value of $\sim$0.5 $\times$ solar
in the regions around SNR N132D.
The mean value of the Ne abundance around 30 Dor is 2.25 times solar, with a mean value of the 90\% confidence range (c.r.) of 1.16. In the regions around N132D, the mean Ne abundance is 1.93 and the mean value of 90 \% c.r. is 0.70. 
In total, a free parameter for the Ne abundance improved the fit in 236 regions, while for 150 regions, the fit was consistent with a Ne abundance of 0.5 times solar.
There are also line transitions of oxygen and magnesium in the energy 
range of 0.4 -- 2.0 keV, in which eROSITA has the highest 
response. However, O and Mg lines are contaminated by background emission 
(solar-wind charge exchange lines and Al K$\alpha$ line at 1.486 keV from the detector, respectively), which were included in the background model as free components. Therefore, the O and Mg abundances are not well constrained.
To understand the possible
origin of Ne, we calculated and plotted the star-formation rate in 
these two areas in LMC (marked with boxes in Fig.\,\ref{imaneabund}) 
based on star-formation rates of \citet{2009AJ....138.1243H}
(Fig.\,\ref{sfr}). In the eastern region (1), 
star formation  has been ongoing in the last $10^6 - 10^7$ years,
producing a large number of young massive stars, which have created
the large and complex HII regions. Most likely, the ISM has been also
enriched by stellar winds and supernovae of these massive stars.

\subsection{Foreground Absorption in the LMC}\label{absorption}

\begin{figure*}
\centering
\includegraphics[width=\textwidth]{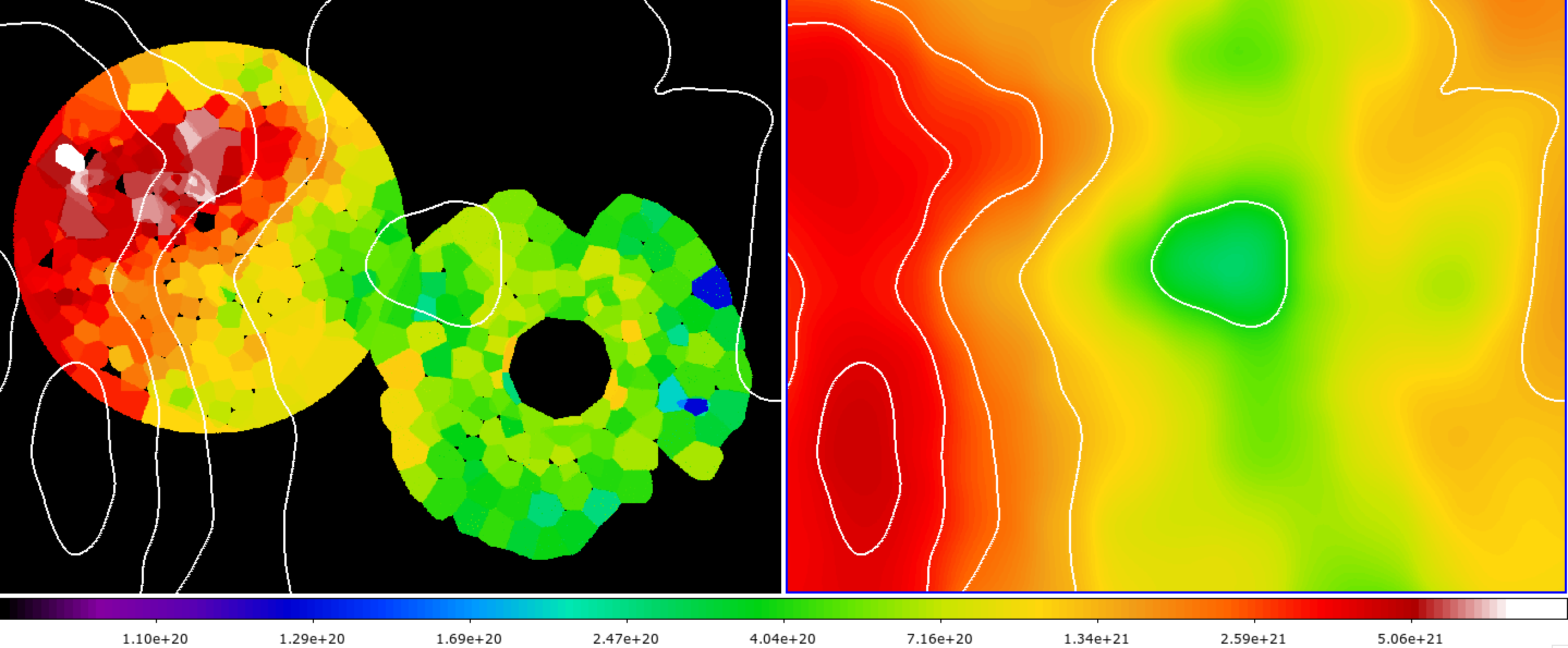}
\caption{\label{imanh}
$N_{\mathrm{H, LMC}}$ in the LMC obtained from the fit of the eROSITA spectra (left) 
vs. $N_{\mathrm{H, cold, LMC}}$  directly calculated from cold \hi\ and dust (right) in cm$^{-2}$.
Both images are shown in log-scale using the same lower and upper cuts. 
Contours of $N_{\mathrm{H, cold, LMC}}$ are plotted on both images.
}
\end{figure*}

The fits have shown that the foreground column density 
$N_{\mathrm{H, LMC}}$
in the LMC is higher around 30~Dor (see Fig.\,\ref{imanh}) than in the west, 
also indicated by the green colour of the X-ray image.
The values obtained from the analysis of the X-ray spectrum
taken with eROSITA is in very good agreement with the 
$N_{\mathrm{H, cold, LMC}}$  in the LMC obtained from the distribution of cold matter
(Sect.\,\ref{nhcold}), as can be seen in Fig.\,\ref{imanh}.
This agreement between the values derived from the X-ray analysis and 
measured directly from the cold medium seen over a large area of $>$1\degr\
is remarkable and shows the advantage of the large field of view
and the high sensitivity of eROSITA in the energy band of 0.2 -- 2.0 keV.

Together with the high normalisation (and thus intrinsic flux) of the
emission in the same regions, the enhanced \nh$_\mathrm{, LMC}$ indicates
that the hot X-ray emitting plasma is 
located behind
a high-density
region. The same stars, which have created the giant \hii\ region
30~Dor and the other \hii\ structures in its surroundings, are
the likely sources of the hot interstellar plasma.
This result corroborates nicely the scenario of the collision of
large HI structures in the LMC for the origin of 30~Dor 
and the star-forming regions south of 30~Dor presented by 
\citet{2017PASJ...69L...5F} and recently confirmed by
\citet{2021A&A...648A..90K}
based on multi-wavelength study
of the southeastern part of the LMC, in particular the X-ray spur. A large HI component, called the
L-component has encountered the HI component in the disk of
the LMC (D-component) in the past, starting roughly at the position where
30~Dor is observed now. The collision has continued south of 30~Dor where young massive stars, and even further to the south, 
a long ridge of dust and CO clouds along with star-forming regions are found next to the X-ray spur.
The L-component is now located in front of the disk of the LMC at
the position of 30~Dor, which corresponds to the eastern part of the 
eROSITA observations that we have analysed, and absorbs the emission 
from 30~Dor as well as from the sources and the hot interstellar plasma around 
it.

\section{Summary}

The first-light observation of eROSITA was pointed at SN 1987A and has provided us with an impressive X-ray view of a large number of sources and complex diffuse emission in the LMC in a large 1\degr-diameter field. This field includes several prominent objects like the giant \hii-region 30 Dor, the non-thermal superbubble 30 Dor C, and bright SNRs and pulsars. At a distance of about 1.5\degr\ from SN 1987A, there is SNR N132D, which is the brightest X-ray SNR in the LMC. As one of the X-ray calibration sources, SNR N132D and its surroundings have also been  observed in the early phase of the eROSITA mission for multiple times. In this paper, we have presented the study of the diffuse X-ray emission in the LMC detected in these early eROSITA observations.

To understand the properties of the hot interstellar plasma and the processes that form it, we have been studying SNRs and superbubbles in the Magellanic Clouds using \xmm\ and \chandra\
\citep[e.g.,][]{2011A&A...528A.136S,2012A&A...547A..19K,2014A&A...567A.136W,2015A&A...573A..73K,2019A&A...621A.138K}. 
Due to a much smaller field of view of these X-ray observatories, our studies so far had to focus on a small number of selected objects. The large field of view of eROSITA and the high sensitivity at energies below 2 keV, combined with the much better spatial and spectral resolution than ROSAT, which was the last X-ray observatory that performed a survey of the entire LMC \citep[for a study of the diffuse emission, see][]{2002A&A...392..103S}, eROSITA is the perfect instrument to study the hot ISM in our neighbour galaxy. Thanks to the early commissioning and calibration observations, we obtained a set of long exposures with eROSITA towards the most interesting regions of the LMC.

We have analysed the data of eleven eROSITA observations by eliminating all point and point-like 
sources
and focusing on the diffuse emission. Using a tessellation algorithm, we divided the data into small regions with a signal-to-noise ratio of $>$100 and $>$50 around SN 1987A and SNR N132D, respectively. The spectra of the diffuse emission in these regions were analysed assuming a combination of a two-component thermal plasma and one non-thermal emission model. Maps of parameter values obtained for the best-fit model have been created and studied. The results of the spectral analysis are:
\begin{itemize}
    \item We detect emission from thermal plasma in the ISM in the LMC in all regions. There is dominant emission from a low-temperature component with $kT$ = 0.2 keV and another lower-brightness thermal component with $kT \approx$ 0.7 keV. The interstellar density and pressure derived from the parameters of the major lower-temperature component are consistent with values measured in the Milky Way.
    \item The emission from the second higher-temperature thermal component is stronger in the environment of 30~Dor, suggesting that the young stellar population has caused recent heating.
    \item In these regions, also the element abundances seem to be enhanced, as indicated by the high Ne abundance. Most likely, the ISM has been enriched by the stellar winds and supernovae of massive stars.
    \item In addition, significant non-thermal emission is confirmed in the superbubble 30~Dor~C. There are indications of the presence of non-thermal emission also east of 30 Dor, which requires further investigation.
    \item The X-ray spectral analysis yields an absorbing column density \nh\ in the LMC, which is surprisingly consistent with the column density derived from the measurements of \hi\ and the gas-to-dust ratio at low energies. This is the first time that the foreground column density has been determined through X-ray spectroscopy over such a large contiguous field and is in agreement with direct measurements from the cold interstellar medium.
    \item We analysed the spectra of the massive stellar cluster RMC 136 and the emission from the Wolf-Rayet stars RMC 139 and RMC 140. The emission is well reproduced by a model consisting of emission from a non-ionisation equilibrium plasma with $kT >$ 1 keV and $\tau = n_e t \approx 10^{11}$ s cm$^{-3}$ and a power-law component with $\Gamma = 1.3$.
    \item Based on eROSITA image and spectroscopy, we confirm SNR J529–7004 as a new SNR in the LMC.
\end{itemize}
As we have been anticipating during the years of preparations for the eROSITA mission, eROSITA is the perfect telescope for studying the ISM at high energies. Currently, eROSITA is carrying out 
a total of eight
all-sky surveys over a period of four years. We will extend the study of the ISM in the LMC based on the eROSITA all-sky survey data. In addition, the eROSITA all-sky survey will also allow us to study the hot phase of the ISM in the SMC.

\begin{acknowledgements}
This work is based on data from eROSITA, the soft X-ray instrument aboard SRG, a joint Russian-German science mission supported by the Russian Space Agency (Roskosmos), in the interests of the Russian Academy of Sciences represented by its Space Research Institute (IKI), and the Deutsches Zentrum f\"ur Luft- und Raumfahrt (DLR). The SRG spacecraft was built by Lavochkin Association (NPOL) and its subcontractors, and is operated by NPOL with support from the Max Planck Institute for Extraterrestrial Physics (MPE).

The development and construction of the eROSITA X-ray instrument was led by MPE, with contributions from the Dr. Karl Remeis Observatory Bamberg \& ECAP (FAU Erlangen-N\"urnberg), the University of Hamburg Observatory, the Leibniz Institute for Astrophysics Potsdam (AIP), and the Institute for Astronomy and Astrophysics of the University of T\"ubingen, with the support of DLR and the Max Planck Society. The Argelander Institute for Astronomy of the University of Bonn and the Ludwig Maximilians Universit\"at Munich also participated in the science preparation for eROSITA.

The eROSITA data shown here were processed using the eSASS/NRTA software system developed by the German eROSITA consortium.

The Australian SKA Pathfinder is part of the Australia Telescope National Facility (ATNF) which is managed by CSIRO. Operation of ASKAP is funded by the Australian Government with support from the National Collaborative Research Infrastructure Strategy. ASKAP uses the resources of the Pawsey Supercomputing Centre. Establishment of ASKAP, the Murchison Radio-astronomy Observatory (MRO) and the Pawsey Supercomputing Centre are initiatives of the Australian Government, with support from the Government of Western Australia and the Science and Industry Endowment Fund. This paper includes archived data obtained through the CSIRO ASKAP Science Data Archive (CASDA). We acknowledge the Wajarri Yamatji as the traditional owners of the Observatory site. 

MCELS was funded through the support of the Dean B. McLaughlin fund at the University of Michigan and through NSF grant 9540747. 

M.S. acknowledges support by the Deutsche Forschungsgemeinschaft through the Heisenberg professor grant SA 2131/12-1.
A.M.B. was supported by the RSF grant 21-72-20020.
\end{acknowledgements}

\bibliographystyle{aa}
\bibliography{../../bibtex/xraytel,../../bibtex/stars,../../bibtex/lmc,../../bibtex/my,../../bibtex/ism,../../bibtex/nearbygal,../../bibtex/doc,../../bibtex/ctb109,../../bibtex/radio,../../bibtex/cosmicray}

\end{document}